\documentclass[11pt,a4paper]{article}
\usepackage{lscape}

    \usepackage{subcaption}
\usepackage[utf8]{inputenc}		
\usepackage[T1]{fontenc}
\usepackage{multirow}
\usepackage{graphicx}
\usepackage{natbib}

\usepackage{booktabs}
\usepackage{rotating}
\usepackage{multirow}
\usepackage{graphicx}
\newcommand{\biblist}{\begin{list}{}
		{\listparindent 0.0cm \leftmargin 0.50cm \itemindent -0.50 cm
			\labelwidth 0 cm \labelsep 0.50 cm
			\usecounter{list}}\clubpenalty4000\widowpenalty4000}
	\newcommand{\ebiblist}{\end{list}}

\usepackage{amsmath,amsthm,amsfonts,amssymb,amscd,amsthm,xspace, dsfont}
\theoremstyle{plain}
\allowdisplaybreaks

\usepackage{latexsym}
\usepackage{amsmath}
\usepackage{amsfonts,amssymb}
\usepackage{multirow}
\usepackage{enumerate}

\newtheorem*{theorem*}{Theorem}

\newtheorem{remark}{Remark}[section]

\newtheorem*{proposition*}{Proposition}

\newtheorem*{result*}{Result}

\newtheorem*{lemma*}{Lemma}

\newcommand{\bx}{\mathbf{x}}

\newcommand{\E}{{\mathbb E}}

\usepackage{hyperref}
\hypersetup{
	colorlinks=true,
	linkcolor=blue,
	filecolor=magenta,
	urlcolor=blue,
	pdftitle={Sharelatex Example},
	bookmarks=true,
	citecolor=blue,
}

\DeclareMathOperator*{\argminA}{arg\,min}
\DeclareMathOperator*{\argmaxA}{arg\,max}

\newcommand{\chapternote}[1]{{%
		\let\thempfn\relax
		\footnotetext[0]{\emph{#1}}
}}

\usepackage[margin=2.8cm]{geometry}

\begin{document}

\title{\bf {\Large Imputation procedures in surveys using nonparametric and machine learning methods: an empirical comparison}}

\author{Mehdi {\sc Dagdoug}$^{(1)}$,  Camelia {\sc Goga}$^{(1)}$ and  David  {\sc Haziza}$^{(2)}$ \\
	(1) Universit\'e de Bourgogne Franche-Comt\'e,\\ Laboratoire de Math\'ematiques de Besan\c con,  Besan\c con, FRANCE \\
	(2) Universit\'e de Montr\'eal, D\'epartement de math\'ematiques et de statistique,\\ Montr\'eal, CANADA
}

\date{\today}
\maketitle
\chapternote{Mehdi Dagdoug's research was supported by grants of the region of Franche-Comt\'e and\\ M\'ediam\'etrie.}

\maketitle

\begin{abstract}
Nonparametric and machine learning methods are flexible methods for obtaining accurate predictions. Nowadays, data sets with a large number of predictors and complex structures are fairly common.  In the presence of item nonresponse, nonparametric and machine learning procedures may thus provide a useful alternative to traditional imputation procedures for deriving a set of imputed values used next for the estimation of study parameters defined as solution of population estimating equation. In this paper, we conduct an extensive empirical investigation that compares a number of imputation procedures in terms of bias and efficiency in a wide variety of settings, including high-dimensional data sets. The results suggest that a number of machine learning procedures perform very well in terms of bias and efficiency.
\end{abstract}


\section{Introduction}

In the last decade, the interest in machine learning methods has been growing in national statistical offices (NSO). These data-driven methods provide flexible tools for obtaining accurate predictions. The increasing availability of data sources (e.g., big data sources and satellite information) provides a rich pool of potential predictors that may be used to obtain predictions at different stages of a survey. These stages include the nonresponse treatment stage (e.g., propensity score weighting and imputation) and the estimation stage (e.g., model-assisted estimation and small area estimation). The imputation stage is the focus of the current paper.

Item nonresponse refers to the presence of missing values for some, but not all, survey variables. Frequent causes of item nonresponse include refusal to answer a sensitive question (e.g., income) and edit failures. The most common way of treating item nonresponse in NSOs is to replace a missing value with a single imputed value, constructed on the basis of a set of $p$ explanatory variables, $\mathbf{X}=(X_1, \ldots, X_p),$ available for both respondents and nonrespondents.  A variety of imputation procedures are available, ranging from simple (e.g., mean, historical and ratio imputation) to more complex (e.g., nonparametric procedures); e.g., see \cite{Chen_Haziza2019} for an overview of imputation procedures in surveys. Every imputation procedure makes some (implicit of explicit) assumptions about the distribution of the variable $Y$ requiring imputation. This set of assumptions is often referred to as an imputation model. At the imputation stage, it is therefore important to identify and include in the model all the appropriate explanatory variables that are predictive of the variable requiring imputation and determine a suitable model describing the relationship between $Y$ and the set of explanatory variables $\mathbf{X}.$

We distinguish parametric imputation procedures from nonparametric imputation procedures. In parametric imputation,  the shape of the relationship between $Y$ and $\mathbf{X}$ is predetermined; e.g., linear and generalized linear regression models. However, point estimators based on parametric imputation procedures may suffer from bias if the functional form is misspecified or if the vector $\mathbf{X}$ fails to include interactions or predictors accounting for curvature. In contrast, with nonparametric methods, the shape of the relationship between $Y$ and $\mathbf{X}$ is left unspecified. These methods have the ability to capture nonlinear trends in the data and tend to be robust to the non-inclusion of interactions or predictors accounting for curvature.

Commonly used nonparametric methods include kernel smoothing, local polynomial regression and spline-based regression models. While these methods provide some robustness against model misspecification, they tend to breakdown when the number predictors is large, a problem known as the curse of dimensionality.  To mitigate this problem, one may employ  additive models \citep{Hastie1986}. However,  when the dimension of $\mathbf{X}$ is very large, these models tend to fail and machine learning methods may provide an interesting alternative. The class of machine learning methods, that includes tree-based models such as random forests and boosting methods, provide more flexible approaches able to adapt to complex non-linear
and non-additive relationships between the survey variable requiring imputation and a set of predictors. These methods may also prove useful in the case of large data sets exhibiting a  large number of observations on a large number of variables.  Many machine learning procedures are relatively computationally efficient and can produce accurate predictions
by offering the user a kind of automatic variable selection that may prove useful in a high-dimensional setting.

However, both a theoretical treatment and an empirical comparison of machine learning imputation procedures in the context of missing survey data are currently lacking. In this paper, we aim to fill the latter gap by conducting an extensive simulation study that investigates the performance of several nonparametric and machine learning procedures in terms of bias and efficiency. To that end, we generated several finite populations with relationships between $Y$ and $\mathbf{X},$  ranging from simple to complex and generated the missing values according to several nonresponse mechanisms. We also considered both a low-dimensional and high dimensional settings. The simulation setup and the models are described in Section \ref{simulation_study}.  We restricted our attention to population totals (Section \ref{simulation_study}) and population quantiles (Section \ref{estimation_quantiles}) as the target parameters.  The following procedures were included in our comparisons: the score method \citep{little1986survey, Haziza2007},  K nearest-neighbour \citep{chen2000nearest}, additive models based on B-spline regression,  regression trees \citep{Breiman_Friedman_Olshen_Stone1984}, random forests \citep{breiman2001random}, tree-based boosting methods \citep{friedman_2001} including XGBoost  \citep{chen2016} and  Bayesian additive regression trees \citep{Chipman2010},  the cubist algorithm \citep{quinlan1992learning, quinlan1993combining} and support vector regression \citep{Vapnik1998, Vapnik2000}. In Section \ref{imputation_methods}, we describe these models and the corresponding imputation procedures.

In recent years, machine learning procedures have received some attention in a survey sampling context.  In the ideal situation of 100\% response, the theoretical properties of model-assisted estimation procedures based on regression trees \citep{mcconville2019automated} and random forests \citep{dagdoug2020model} have been recently established.  \cite{dagdoug_goga_haziza_RF_NR} studied the theoretical properties  of point and variance estimators based on random forests in a context of imputation for item nonresponse and data integration; see also \cite{tipton2013properties, de2018sample} for applications of random forests in surveys.   A number of empirical investigations  have been conducted to assess the performance of machine learning procedures in a context of propensity score estimation for unit nonresponse; e.g.,  \cite{Lohr_Hsu_Montaquila}, \cite{Gelein_B.2017} and \cite{Kern_Klausch_Kreuter_2019}.

The machine learning procedures described in Section \ref{imputation_methods} slightly differ from their traditional implementation because of the inclusion of the sampling weights in the construction of imputed values. However, it should be noted that most of the machine learning software packages for obtaining predicted values assume simple random sampling and cannot handle unequal weights. Modifying machine learning algorithms to account for unequal weights may prove challenging.
When the design features  (e.g., sampling weights, stratum indicators, etc.)  are related to the survey variable requiring imputation, failing to incorporate them in the models may lead to biased estimators.  To cope with this issue, we suggest to include all the appropriate design variables in the specification of the model. Standard machine learning software packages may then be safely used for creating a set of imputed values. In Section \ref{simulation_study}, we use Poisson sampling with inclusion probabilities proportional to a size variable $X$ to select repeated samples from the finite population. The size variable $X$ being related to the variable requiring imputation, including the $X$-variable in the specified models led to satisfactory results.

\section{Preliminaries}\label{preliminaries}
Consider a finite population $U = \left\{1, 2, ..., N\right\}$ of size $N$. Let $Y$ denote a survey variable and $y_i$ be the $y$-values attached to unit $i,$ $i=1, \cdots, N.$ We are interested in estimating  (i) the finite population total of the $y$-values, $t_y = \sum_{i \in U} y_i$ and (ii) the finite population quantile of order $\gamma$ defined as $\mathcal{Q}_{\gamma} := \inf  \left\{t \in \mathbb{R} ;  F_N(t) \geqslant \gamma \right\},$ where
$$F_N(t) = \sum_{i \in U } \mathds{1} \left(y_i \leqslant t\right)/N$$
denotes the finite population distribution function.

From $U,$ we select a sample $S$, of size $n$, according to a sampling design $\mathcal{P} \left(S=s\right)$ with first-order inclusion probabilities $\pi_i=Pr(i\in S)$.


A complete data estimator of $t_y$ is the well-known Horvitz-Thompson estimator
\begin{equation}\label{HT}
\widehat{t}_{\pi} = \sum_{i \in S} \dfrac{y_i}{\pi_i},
\end{equation}
which is design-unbiased for $t_y$ provided that $\pi_i>0$ for all $i \in U$. A complete data estimator of the finite population quantile $\mathcal{Q}_{\gamma}$ is given by
 \begin{equation}\label{comp_quantile}
 	\widehat{\mathcal{Q}}_{\gamma}:= \inf  \left\{t \in \mathbb{R} ; \widehat{F}(t) \geqslant \gamma \right\},
 \end{equation}
 where
 \begin{equation} \label{quantileS}
 	\widehat{F}(t) = \dfrac{1}{\widehat{N}}\sum_{i \in S } \dfrac{\mathds{1} \left(y_i \leqslant t\right)}{\pi_i}
 \end{equation}
 with $\widehat{N}=\sum_{i\in S}1/\pi_i$ denoting the Horvitz-Thompson estimator of the population size $N$. Under mild regularity conditions \citep{wang_opsomer_2011}, the complete data estimator  $\widehat{\mathcal{Q}}_{\gamma}$ is design-consistent for $\mathcal{Q}_{\gamma}.$


In practice, the $Y$-variable may be prone to missing values. Let $r_i$ be a response indicator such that $r_i=1$ if $y_i$ is observed and $r_i = 0,$ otherwise.  Let $S_r = \left\{i \in S; \ r_i = 1 \right\}$ denote the set of respondents, of size $n_r,$ and $S_m = \left\{i \in S; \  r_i = 0 \right\}$ the set of nonrespondents, of size $n_m,$ such that $S_r \cup S_m = S$ and $n_r + n_m=n$. Available to the imputer is the data $(y_i, \mathbf{x}_i)$ for $i \in S_r$ as well as the values of the vector $\mathbf{x}_i$ for $i \in S_m.$

Let $\widehat{y}_i$ be the imputed value used to replace the missing value $y_i$ and
$$\widetilde{y}_i=r_i y_i + \left(1 - r_i\right)\widehat{y}_i$$ be the $i$th value of the $Y$-variable after imputation. Point estimators of $t_y$ and  ${\mathcal{Q}}_{\gamma}$ after imputation, often referred to as imputed estimators,   are readily obtained from the complete data estimators (\ref{HT}) and (\ref{comp_quantile}) by replacing $y_i$ with $\widetilde{y}_i$. This leads to
\begin{equation} \label{imputed}
\widehat{t}_{imp}
=\sum_{i\in S}\frac{\widetilde y_i}{\pi_i}
\end{equation}
and
\begin{equation} \label{imputed_quantile}
\widehat{\mathcal{Q}}_{\gamma, imp}= \inf  \left\{t \in \mathbb{R} ; \widehat{F}_{imp}(t) \geqslant \gamma \right\},
\end{equation}
where
\begin{equation} \label{quantileSr}
	\widehat{F}_{imp}(t) = \dfrac{1}{\widehat{N}} \sum_{i \in S } \dfrac{\mathds{1} \left(\widetilde y_i \leqslant t\right)}{\pi_i}
\end{equation}
denotes the imputed estimator of $F_N(t)$.

\begin{remark}
The population total $t_y,$  the distribution function $F_N(t)$ and the quantile of order $\gamma,$ $\mathcal{Q}_{\gamma},$ may all be obtained as the solution of the following census estimating equation  \citep{binder_1983, Chen_Haziza2019}:
	\begin{eqnarray}
	U_N(\theta_N)=\sum_{i\in U}u(y_i; \theta_N)=0,\label{est_equat_U}
	\end{eqnarray}
	where $\theta_N$ is a generic notation denoting a finite population parameter and  $u(y_i; \theta)$ is a function of $ \theta_N$. We assume that a solution to (\ref{est_equat_U})  exists and is unique. For instance, the population total $t_y$ can be obtained as a solution of  (\ref{est_equat_U}) with $u(y_i;  \theta_N)=y_i-n^{-1}\pi_i\theta_N$;   the finite population distribution function $F_N(t)$ can be obtained as a solution of (\ref{est_equat_U}) with $u(y_i;\theta_N)=\mathds{1} \left(y_i \leqslant t\right)-\theta_N.$ Finally, the quantile $\mathcal{Q}_{\gamma}$ of order $\gamma$ can be obtained as a solution of (\ref{est_equat_U}) with $u(y_i;\theta_N)=\mathds{1} \left(y_i \leqslant \theta_N\right)-\gamma.$ Other finite population parameters can be obtained as a solution of (\ref{est_equat_U}); e.g., see  \cite{Chen_Haziza2019}.
The imputed estimators $\widehat{t}_{imp},$ $\widehat{\mathcal{Q}}_{\gamma, imp}$ and  $\widehat{F}_{imp}(t)$ given respectively by (\ref{imputed})-(\ref{quantileSr}) can be obtained by solving the following sample estimating equation:
\begin{eqnarray*}
\widehat{U}_{imp}(\widehat{\theta}_{imp})=\sum_{i\in S}\frac{1}{\pi_i}u(\widetilde{y}_i; \widehat{\theta}_{imp})=0,\label{est_equat_s_imp}
\end{eqnarray*}
where $\widehat{\theta}_{imp}$ denotes an imputed estimator of $\theta_N.$
\end{remark}

To construct the imputed values $\widehat{y}_i$, we postulate the following imputation model $\xi$:
\begin{align}\label{first_mom}
&\E_\xi (y_i \rvert \bx_i) = f(\bx_i), \\
&\mathbb{V}_\xi \left(y_i \rvert \bx_i\right) = \sigma_i^2, \nonumber \\
&\mathbb{C}ov_\xi \left(y_i, y_j \rvert \bx_i, \bx_j \right) = 0 \quad \mbox{for } \qquad i \neq j\nonumber,
\end{align}
where $f$ is an unknown function. Often, the variance structure $\sigma^2_i$ is assumed to have the form $ \sigma^2_i= \sigma^2 a_i,$ where $a_i>0$  is a known coefficient attached to unit $i$ and  $\sigma^2$ is an unknown parameter.

We assume that the data are Missing At Random \citep{Rubin1976}:
\begin{equation} \label{mar}
	f(y_i \rvert \bx_i, r_i =1) = f(y_i \rvert \bx_i, r_i =0).
\end{equation}
That is, we assume that the distribution of $Y$ given $\mathbf{x}$ is the same for both respondents and nonrespondents. If Condition (\ref{mar}) holds, the imputed values can be safely generated from $	f(y_i \rvert \bx_i, r_i =1),$ which can be estimated from the observed data. In the context of imputation, the properties of point estimators are evaluated with respect to the joint distribution induced by the imputation, the sampling design and the unknown nonresponse mechanism. This framework is often referred to as the $\xi pq$-framework \citep{Chen_Haziza2019}. Note that our simulation setup in Section \ref{simulation_study} is consistent with the $\xi pq$-framework as the simulation process involves (i) generating repeated finite populations; (ii) selecting a sample from each of population and (iii) generating a set of response indicators in each sample.

Deterministic imputation consists of replacing the missing $y_i$ by   $\widehat{y}_i=\widehat{f}(\bx_i),$ where $\widehat{f}$ is an estimator of the unknown regression function $f$ based on the responding units $i \in S_r$.
However, deterministic imputation methods tend to distort the distribution of the survey variable $Y$ requiring imputation, potentially leading to biased estimators of quantiles \citep{haziza_2009, Chen_Haziza2019}. To cope with this issue, one can recourse to random imputation that consists of adding an appropriate amount of random noise to the deterministic value $\widehat{f}(\bx_i)$. More specifically, let $e_j :=\widehat{\sigma}_j^{-1}\{y_j -  \widehat{f}(\bx_j)\}$ for $j \in S_r$, where $\widehat{\sigma}_j$ of an estimator of $\sigma_j$ (see Remark \ref{Rem1} below).  We define the standardized residual  
$$
\widetilde{e}_j = e_j - \dfrac{\sum_{\ell \in S_r} w_\ell e_\ell }{\sum_{\ell\in S_r} w_\ell}, \quad  j\in S_r.
$$
In the case of random imputation, the missing $y_i$ is replaced by
\begin{equation}\label{RI}
\widehat{y}_i = \widehat{f}(\bx_i) + \widehat{\sigma}_i \widehat{e}_i,
\end{equation}
where $\widehat{e}_i$ is selected at random from the set of standardized residuals $\{\widetilde{e}_j\}_{j \in S_r}$ with probability $ w_j/ \sum_{ \ell \in S_r} w_\ell$.

\begin{remark}\label{Rem1}
To obtain an estimator $\widehat{\sigma}_i$ of $\sigma_i,$ one can postulate a model  $\E(\epsilon_i^2 \mid \mathbf{x}_i)=m(\mathbf{x}_i),$ where $m$ is an unknown function. An estimator $\widehat{\sigma}^2_i$ of $\sigma^2_i$ is obtained by fitting a parametric or a nonparametric procedure with the square residuals $e_i^2$ as the response and $\mathbf{x}_i$ as the set of predictors.
	\end{remark}

In Section \ref{imputation_methods}, except for the parametric imputation procedure discussed in Section \ref{linearreg}, all the other procedures (Section \ref{score_method}-\ref{SVR}) are nonparametric. In Section \ref{simulation_study}, these procedures are compared empirically in terms of bias and efficiency under a variety of settings.

\section{A description of imputation methods}\label{imputation_methods}

\subsection{Parametric regression imputation}
\label{linearreg}

Parametric regression assumes that the first moment (\ref{first_mom}) is given by
\begin{equation}\label{specified}
\E_\xi (y_i\rvert \bx_i)  = f (\mathbf{x}_i, \boldsymbol{\beta}),
\end{equation}
where  $\boldsymbol{\beta}$ is a vector of coefficients to be estimated and $f(\cdot)$ is a  predetermined function. An estimator $\widehat{\boldsymbol{\beta}}$ of $\boldsymbol{\beta}$ is obtained by solving the following estimating equations based on the responding units:
\begin{equation}
\sum_{ i \in S_r} \dfrac{w_i}{\sigma_i^2} \left\{y_i - f (\mathbf{x}_i, \boldsymbol{\beta}) \right\} \dfrac{\partial f (\mathbf{x}_i, \boldsymbol{\beta}) }{\partial \boldsymbol{\beta}} = 0,\label{eq_est_beta}
\end{equation}
where $w_i > 0$ is a weight attached to element $i$. Common choices for $w_i$ include $w_i=1$ and $w_i = \pi_i^{-1}$  \citep{Chen_Haziza2019}. The imputed value $\widehat{y}_i$ under deterministic parametric regression imputation is given by
\begin{equation}\label{SPR}
\widehat{y}_i = f(\bx_i, \widehat{\boldsymbol{\beta}}), \quad i \in S_m.
\end{equation}
A special case of (\ref{SPR}) is $f (\mathbf{x}_i, \boldsymbol{\beta})=\bx_i^\top\boldsymbol{\beta},$ which corresponds to the customary linear regression model. In this case, the imputed value (\ref{SPR}) reduces to
\begin{equation}\label{imp_lin_mod}
\widehat{y}_i = \bx_i^\top \widehat{\boldsymbol{\beta}},\quad i \in S_m,
\end{equation}
where
\begin{equation}\label{regLin}
\widehat{\boldsymbol{\beta}} = \left( \sum_{j \in S_r}w_j\sigma^{-2}_j \bx_j \bx_j^\top\right)^{-1}  \sum_{j \in S_r}w_j\sigma^{-2}_j \bx_j y_j.
\end{equation}
The imputed value $\widehat{y}_i$ given by (\ref{imp_lin_mod}) can be written as a weighted sum of the respondent $y$-values:
\begin{eqnarray}\label{LR_weighted_form}
\widehat{y}_i=\sum_{j\in S_r}w'_{ij}y_j, \quad i\in S_m,
\end{eqnarray}
where $w'_{ij}= \bx_i^\top\left( \sum_{j' \in S_r}w_{j'}\sigma^{-2}_{j'} \bx_{j'} \bx_{j'}^\top\right)^{-1} w_j  \sigma^{-2}_j\bx_j. $ If the intercept is among the $X$-variables, then $\sum_{j\in S_r}w'_{ij}=1$ for all $i\in S_m.$  A random counterpart of (\ref{SPR}) is given by (\ref{RI}).

Another important special case of (\ref{SPR}) is the logistic regression model,  $$f(\mathbf{x}_i, \boldsymbol{\beta})=\exp(\bx_i^\top\boldsymbol{\beta})/(1+\exp(\bx_i^\top\boldsymbol{\beta})),$$
which can be used for modeling binary variables. An estimator of $\boldsymbol{\beta}$ is obtained by solving (\ref{eq_est_beta}), which requires a numerical algorithm such as the Newton-Raphson procedure. To eliminate the possibility of an impossible imputed value,  a missing value to a $0-1$ variable is typically imputed by $\widehat{y}_i,$ where $\widehat{y}_i$ is a realization of a Bernoulli variable with parameter $f(\mathbf{x}_i, \widehat{\boldsymbol{\beta}}).$

Under deterministic or random parametric regression imputation, the imputed estimator  $\widehat t_{imp}$ is consistent for $t_y$ provided that the first moment of the imputation model  (\ref{first_mom}) is correctly specified. However, this type of imputation may lead to biased estimators of quantiles. In contrast, the use of a random parametric regression imputation procedure tend to preserve the distribution of the variable requiring imputation, leading to valid estimators; see \cite{Chen_Haziza2019} for a discussion.
%

\subsection{Imputation classes : the score method}\label{score_method}
The score method  \citep{little1986survey, Haziza2007}  consists of partitioning the sample $S$ into $H$ (say) imputation classes and imputing the missing values within each class independently from one class to another. It can be implemented as follows:
\begin{enumerate}[Step 1:]
	\item  For all $i \in S$, compute the preliminary values $\widehat{y}_i^{LR} = \mathbf{x}_i^{\top} \widehat{\boldsymbol{\beta}},$ where $\widehat{\boldsymbol{\beta}}$ is given by (\ref{regLin}).
	\item Compute the empirical quantiles $q_1, q_2, \ldots, q_{H-1}$ of order $1/H, 2/H, \ldots, (H-1)/H$ of the $\widehat{y}^{LR}$-values.
	\item Split the sample $S$ into $H$ classes, $C_1, \ldots, C_h, \ldots, C_H,$ such that
	$$C_h = \left\{i \in S: \widehat{y}_i^{LR} \in [q_{h-1} ; q_h )\right\}, \quad h=1, \ldots, H,$$
	with $q_0=- \infty$ and $q_H=+ \infty$.
\end{enumerate}
It is common practice to use either mean imputation or random hot-deck imputation within classes. For mean imputation, the imputed value for missing $y_i$ in the $h$th imputation class is given by
\begin{equation*}
\widehat{y}_i = \frac{\sum_{j \in S_r\cap C_h } w_j y_j}{\sum_{j \in S_r\cap C_h} w_j}=\sum_{j\in S_r\cap C_h}w'_{ij}y_j, \quad i\in S_m\cap C_h,
\end{equation*}
where $w'_{ij}=w_j/\sum_{j' \in S_r\cap C_h} w_{j'}$ are the same for all $i\in S_m\cap C_h$ and $\sum_{j \in S_r\cap C_h} w'_{ij}=1$ for all $i\in S_m\cap C_h.$
For random hot-deck imputation, the imputed value is given by $\widehat{y}_i = y_{j},$ where the donor $j \in S_r\cap C_h $ is selected at random from the set of donors belonging to the $h$th imputation class with probability ${w_j}/{\sum_{j'\in S_r\cap C_h} w_{j'}}.$ Note that random hot-deck imputation within classes can be viewed as mean imputation within classes with added residuals.

\subsection{$K$-nearest neighbours imputation}\label{KNN}
$K$-nearest neighbour ($K$NN) imputation is one of the simplest and widely used nonparametric imputation procedures. No explicit assumption is made about the regression function $f$ relating  $Y$ and $\mathbf{X}$.  $K$NN imputation consists of replacing the missing value of a recipient by the weighted average of the $y$-values of its $K$ closest respondents in terms of the $X$-variables.


Nearest-neighbour (NN) imputation corresponds to the limiting case of $K$NN obtained with $K=1$.  NN is a donor imputation belonging to the class of hot-deck procedures \citep{chen2000nearest}  since a missing value is replaced by an actual respondent $y$-value from the same file. NN imputation is especially useful for imputing categorical or discrete $Y$-variables; e.g., see  \cite{chen2000nearest},  \cite{beaumont_bocci_2009} and \cite{yang2019nearest}.

Let $\mathcal N_K(i)$ be the set of $K$ responding units closest to $\bx_i.$
Any distance function in $\mathbb R^p$ may be used to measure the closeness between two vectors $\bx_i$ and $\bx_j$. In the simulation study presented in Section \ref{simulation_study}, we used the customary Euclidean distance. The $K$NN imputed value for missing $y_i$ is given by
\begin{equation*}
\widehat{y}_i = \dfrac{\sum_{j \in \mathcal N_K(i) \cap S_r} w_j y_j}{\sum_{j \in \mathcal N_K(i) \cap S_r} w_j }, \quad i\in S_m.
\end{equation*}
The imputed value $\widehat{y}_i$ obtained with $K$NN can be written as a weighted sum of the respondent $y$-values:
$$
\widehat{y}_i =\sum_{j\in S_r}w'_{ij}y_j,\quad i\in S_m,
$$
where $w'_{ij}=w_j\mathds{1}(j\in \mathcal N_K(i))/\sum_{j'\in \mathcal N_K(i) \cap S_r}w_{j'}$ for $j\in  S_r$ with $\sum_{j\in S_r}w'_{ij}=1. $  $K$NN imputation is a locally weighted procedure since the respondents $j$ lying not close enough to unit $i$ with respect to the $X$-variables are assigned a weight equal to 0; i.e.,  $w'_{ij}=0. $  The indicator function in the expression of $w'_{ij}$ can be replaced by a one-dimensional continuous kernel smoother $\mathcal K_h,$ whose role is to control the size of the weight through a tuning parameter $h:$ the units $j$ lying farther from unit $i$ will be assigned a smaller weight than units lying close to it \citep{hastie_tibshirani_friedman_2011}. 

%

The imputed estimator under $K$NN imputation tends to be inefficient when the dimension $p$ of $\bx$ is large. Indeed, as $p$ increases, it becomes more difficult to find  enough respondents around the point at which we aim to make a prediction. This phenomenon is known as the curse of dimensionality  \cite[Chap. 1]{hastie_tibshirani_friedman_2011} for a more in-depth discussion ok the $K$NN procedure. Also, it suffers from a model bias which is of order $(K/n)^{1/p}.$
Nearest-neigbour imputation for missing survey data has been considered in \cite{chen2000nearest},  \cite{beaumont_bocci_2009} and \cite{yang2019nearest}.

%

\subsection{B-splines and additive model nonparametric regression}\label{Bsplines}

Spline regression is a flexible nonparametric method for fitting non-linear functions $f(\cdot)$. It can be viewed as a simple extension of linear models.  For simplicity, we start with a univariate $X$-variable supported on the interval $[0;1]. $ 
A spline function of order $v$ with $\kappa$ equidistant interior knots, $0 = \xi_0 < \xi_1 < ...< \xi_{\kappa} < \xi_{\kappa+1}= 1,$ is a piecewise polynomial of degree $v-1$ between knots and smoothly connected at the knots. 
These spline functions span a linear space of dimension of $q=v+\kappa$ with a basis function  given by the $B$-splines functions:
$$B_{\ell}(x)=(\xi_{\ell}- \xi_{\ell-v})\sum_{l=0}^v (\xi_{\ell-l}-x)_+^{v-1}/\Pi_{r=0, r\neq l}^v(\xi_{\ell-l}-\xi_{\ell-r}), \quad \ell=1, \ldots, q,$$
where $(\xi_{\ell-l}-x)_+^{v-1}=(\xi_{\ell-l}-x)^{v-1}$ if $\xi_{\ell-l}\geq x$ and equal to zero, otherwise; see \citep{schumaker_1981, dierckx_1993}.
The $B$-spline basis is appealing because the basis functions are strictly local: each function $B_{\ell}(\cdot)$ has the knots  $\xi_{\ell-v}, \ldots, \xi_{\ell}$ with $\xi_{r}=\xi_{\min (\max(r,0), \kappa+1)}$ for $r=\ell-v,\ldots, \ell$ \citep{zhou_shen_wolfe_1998}, which means that its support consists of a small, fixed, finite number of intervals between knots.
 The unknown function $f(\cdot)$ is then approximated by $\widehat f(\cdot),$  a linear combination of basis functions $\{B_{\ell}\}_{\ell=1}^{q}$ with coefficients determined by a least squares criterion computed on the data $(y_i, x_i)_{i\in S_r}$ \citep{gogahaziza_splines2019}. The missing value $y_i$ is then imputed by $\widehat y_i=\widehat f(x_i),$ where
\begin{equation}\label{Bspline_imp}
\widehat f(x_i) = \sum_{\ell = 1}^{q} \widehat{\beta}_{\ell} B_\ell(x_i)=\mathbf{b}^{\top}_i\widehat{\boldsymbol{\beta}}, \qquad x_i \in [0 ; 1],
\end{equation}
with $\mathbf{b}_i=(B_{\ell}(x_i))_{\ell=1}^{q}$ denoting the vector of $B$-spline basis functions, and $\widehat{\boldsymbol{\beta}}=(\widehat{\beta}_{\ell})_{\ell=1}^{q}$ minimizes
\begin{eqnarray}
\widehat{\boldsymbol{\beta}}&=&\mbox{arg}\min_{\boldsymbol{\beta}\in \mathbf{R}^{q}}\sum_{j \in S_r}w_j\left(y_j - \sum_{\ell = 1}^{q} \beta_\ell B_\ell (x_j)\right)^2=  \left( \sum_{j \in S_r}w_j\mathbf{b}_j \mathbf{b}_j^\top\right)^{-1}  \sum_{j \in S_r}w_j\mathbf{b}_j y_j;\label{opt_Bsplines}
\end{eqnarray}
see \cite{gogahaziza_splines2019}. The expression of $\widehat{\boldsymbol{\beta}}$ is similar to that obtained with linear regression imputation given by (\ref{regLin}) but unlike (\ref{regLin}), the estimator (\ref{opt_Bsplines}) uses the $B$-spline functions $B_1, \ldots, B_{q},$ whose number can vary as a function of the number of knots $\kappa$  and the order $v$ of the $B$-spline functions. The degree $v$ of the piecewise polynomial does not seem to have a great impact on the model fits if a large enough number of interior knots is used \citep{ruppert_wand_carroll_2003}. This is why quadratic or cubic splines are mostly used in practice and an adequate number of interior knots will allow to obtain flexible fits that capture local non-linear trends in the data. Knots are usually placed at the $X$-quantiles and their number may have a great effect on the model fits: a large value of $\kappa$ will lead to overfitting, in which case a penalization criterion may be used in  (\ref{opt_Bsplines}), while a small value of  $\kappa$ may lead to underfitting. \cite{ruppert_wand_carroll_2003}  give a practical rule for choosing the number $\kappa$ of interior knots :
$$
\kappa=\min\left(\frac{1}{4}\times \mbox{number of unique } x_i, 35\right).
$$

The imputed value (\ref{Bspline_imp}) with $B$-spline regression can be also written as a weighted sum of the respondent $y$-values similar to (\ref{LR_weighted_form}), $\widehat{y}_i=\sum_{j\in S_r}w'_{ij}y_j$ for all $i\in S_m$
with weights now given by $w'_{ij}=\mathbf{b}_i^\top\left( \sum_{j' \in S_r}w_{j'} \mathbf{b}_{j'} \mathbf{b}_{j'}^\top\right)^{-1}  w_j  \mathbf{b}_j. $ These weights do not depend on the $y$-values as in linear regression imputation  and $\sum_{j\in S_r}{w'_{ij}}=1$ since $\sum_{j=1}^qB_j(x)=1$ for all $x\in [0;1].$ Unlike linear regression imputation, the weights $w'_{ij}$ are now local due to the $B$-spline functions ensuring more flexibility to model local nonlinear trends in the data.

We now turn to the multivariate case. For ease of presentation, we confine to the case of two predictors, $X_1$ and $X_2.$ Additive models provide a simple way to model nonlinear trend in the data \citep{Hastie1986} and extend the standard linear model by allowing non-linear functions between the response variable $Y$ and each of the explanatory variables, while maintaining additivity. In the case of two predictors, the relationship between $Y$ and $X_1, X_2$ is expressed as a linear combination of unknown smooth functions $f_1$ and $f_2$:
\begin{equation}
y_i= \alpha+f_1 (x_{i1}) + f_2 (x_{i2})+\epsilon_i,\label{model_additif}
\end{equation}
where the $\epsilon_i$'s are independent errors with mean equal to zero.  The model (\ref{model_additif}) is restricted to be additive and does not account for the potential  interactions among the predictors. Accounting for interactions between $X_1$ and $X_2$ would require the additional predictor $X_1X_2$ to be included in the model, leading to
\begin{eqnarray*}
y=f_1(x_1)+f_2(x_2)+f_3(x_1,x_2)+\xi, \label{model_additif2}
\end{eqnarray*}
where $f_3$ is a low-dimensional interaction function fitted by using two-dimensional smoothers, such as local regression or two-dimensional splines. This is beyond the scope of this article. When the number of predictors is large, the number of potential interactions may be considerable, making the implementation of this procedure challenging. In such situations, random forests and boosting, discussed in sections \ref{RF} and \ref{boosting}, provide more flexible approaches. But, as pointed out by \cite{hastie_tibshirani_2015}, additive models provide a useful compromise between linear and fully nonparametric models.

The unknown functions $f_1$ and $f_2$ in (\ref{model_additif}) can be estimated by using two $B$-spline basis $\mathcal B_1=\{B_{11}, \ldots, B_{1q_1}\}$ and $\mathcal B_2=\{B_{21}, \ldots, B_{2q_2}\},$ which leads to $\widehat f_1(x_{i1})=\sum_{\ell = 1}^{q_1} \widehat{\beta}_{1\ell} B_{1\ell}(x_{i1})$ and $\widehat f_2(x_{i2})=\sum_{\ell = 1}^{q_2} \widehat{\beta}_{2\ell} B_{2\ell}(x_{i2}),$ where $\widehat{\beta}_{1\ell}$ and $\widehat{\beta}_{2\ell}$ are determined, as before,  by a least square criterion. To ensure the identifiability of $\alpha,$ additional constraints such as  $\sum_{i=1}^{n_r}\widehat{f}_{1} (x_{i1})=\sum_{i=1}^{n_r}\widehat{f}_{2} (x_{i2})=0$ are usually imposed. With these constraints, the estimators $(\widehat \alpha, \widehat{\boldsymbol{\beta}}_{1},\widehat{\boldsymbol{\beta}}_{2})$ are simply obtained as a regression coefficient estimator, for $\widehat{\boldsymbol{\beta}}_{1}=(\widehat{\beta}_{1\ell})_{\ell=1}^{q_1}$ and $\widehat{\boldsymbol{\beta}}_{2}=(\widehat{\beta}_{2\ell})_{\ell=1}^{q_2}.$
 The imputed value for missing $y_i$ is given by
\begin{equation}\label{AM_imp}
\widehat{y}_i= \widehat{\alpha}+  \widehat{f}_1 (x_{i1})+\widehat{f}_2 (x_{i2}), \quad i\in S_m.
\end{equation}
In practice, a backfitting algorithm is used to compute $f_1(\cdot)$ and $f_2(\cdot)$ iteratively \citep{hastie_tibshirani_friedman_2011}. However, when the number $p$ of explanatory variables is large, the algorithm may not converge and additive models tend to breakdown.  Finally, random versions of (\ref{Bspline_imp}) and (\ref{AM_imp}) are obtained by adding random residuals as in (\ref{RI}).
\subsection{Regression trees}\label{CART}

Regression trees through the CART algorithm have been initially suggested by \citet{breiman1984classification}. Tree-based methods are simple to use in practice for both continuous and categorical variables and useful for interpretation. They form a class of algorithms which recursively split the $p$-dimensional predictor space, the set of possible values for the $X$-variables, into distinct and non-overlapping regions of $\mathbb R^p$. The prediction $\widehat{f}_{tree}(\mathbf{x}_i)$ at point $\mathbf{x}_i$ corresponds to the average of the respondent $y$-values falling in the same region as unit $i$. When the number of $X$-variables is not too large, the splitting algorithm is quite fast, otherwise it may be time-consuming.


Following
\cite{CreelD.&KrotkiK.2006}, we slightly adapt the original CART algorithm as well as the estimation procedure of $f(\cdot)$.
The CART algorithm recursively searches for the splitting variable and the splitting position (i.e., the coordinates on the predictor space where to split) leading to the greatest possible reduction in the residual mean of squares before and after splitting.
More specifically, let $A$ be a region or node and let $\#(A)$  the number of units belonging to $A. $ A split in $A$ consists of finding a pair $(\ell,z),$ where $\ell$ is the variable coordinates taking value between $1$ and $p,$ and $z$ is the position of the split
along the $\ell$th coordinate, within the limits of $A. $ Let $\mathcal{C}_A$ be the set of all possible pairs $(\ell,z)$ in $A$.
The splitting process is performed by searching for the best split $(\ell^*,z^*)$ in the sense that
\begin{eqnarray} \label{opt1}
\left(\ell^*, z^*\right) &= &\argmaxA_{\left(\ell,z\right) \in \mathcal{C}_A} L(\ell,z)\quad \\\mbox{with}\nonumber\\
\hspace{-0.5cm} L(\ell,z) &=& \dfrac{1}{\#(A)} \sum_{i \in S_r} \mathds{1}(\mathbf{x}_i \in A) \left\{ \left(y_i - \bar{y}_A\right)^2 -  \left(y_i -\bar{y}_{A_{L}} \mathds{1}(X_{i\ell} < z) -\bar{y}_{A_{R}}\mathds{1}(X_{i\ell} \geqslant z) \right)^2 \right\},\nonumber\\
\end{eqnarray}
where $X_{ij}$ is the measure of $j$th variable $X_j$ for the $i$th individual, $A_{L} = \left\{ \textbf{X} \in A ; \textbf{X}_{\ell} < z \right\}$, $A_{R} = \left\{ \textbf{X} \in A ; \textbf{X}_{\ell} \geqslant z \right\}$ and $\textbf{X}_{\ell}$ the $\ell$th coordinate of $X;$ $\bar{y}_A$  is the average of $y_i$ for those units $i$ such that $\bx_i \in A$. In (\ref{opt1}), $\mathds{1}(\bx_i\in A)=1$ if $\bx_i\in A,$ and $\mathds{1}(\bx_i\in A)=0,$ otherwise.
From (\ref{opt1}), the best split $\left(\ell^*, z^*\right)$ is the one that produces a tree with the smallest residuals sum of squares \citep[Chap. 8]{hastie_tibshirani_2015}; that is,  we seek $\left(\ell^*, z^*\right)$ that minimizes
$$
\left(\ell^*, z^*\right)=\mbox{arg}\min_{\left(\ell,z\right) \in \mathcal{C}_A}\left\{ \sum_{i \in S_r:\mathbf{x}_i \in A} \left(y_i -\bar{y}_{A_{L}}\right)^2 \mathds{1}(X_{i\ell} < z) + \sum_{i \in S_r:\mathbf{x}_i \in A} \left(y_i-\bar{y}_{A_{R}}\right)^2\mathds{1}(X_{i\ell} \geqslant z)\right\}.
$$
The missing $y_i$ is replaced by $\widehat{y}_i =\widehat f_{tree}(\bx_i),$ which corresponds to the weighted average of the respondent $y$-values falling into the same region as $i \in S_m:$
\begin{equation}
\widehat{y}_i = \sum_{j \in S_r} \dfrac{w_j \mathds{1}(\bx_j\in A(\bx_i))}{\sum_{j'\in S_r} w_{j'}\mathds{1}(\bx_{j'}\in A(\bx_{i}))}y_j, \quad i\in S_m,\label{hat_f_tree}
\end{equation}
where $A(\bx_i)$ is the region from $\mathbb R^p$  containing the point $\bx_i.$  With tree-based methods, the imputed value $\widehat{y}_i$ can also be expressed as
\begin{equation}\label{imp_tree}
\widehat{y}_i = \sum_{j \in S_r}w'_{ij} y_j, \quad i\in S_m,
\end{equation}
where $w'_{ij}=w_j \mathds{1}(\bx_j\in A(\bx_i))/\sum_{j'\in S_r} w_{j'}\mathds{1}(\bx_{j'}\in A(\bx_i))$ with $\sum_{j\in S_r}w'_{ij}=1. $
 With regression trees and tree-based methods in general, the non-overlapping $A$-regions obtained by means of the CART algorithm depend on the respondent data $\{(y_i, \bx_i)\}_{i\in S_r};$ i.e., the same set of $X$-variables with a different set of respondents will lead to different  non-overlapping $A$-regions. The resulting imputed estimator is similar to a post-stratified estimator based on adaptative post-strata.

Regression trees are simple to interpret and often exhibit a small model bias. However, they tend to overfit the data if each $A$-region contains too few elements. To cope with this issue, regression trees may be pruned, meaning that superfluous splits (with respect to a penalized version of (\ref{opt1})) are removed from the tree. Pruning a regression tree tends to reduce its model variance at the expense of increasing the model bias;  see \cite{hastie_tibshirani_friedman_2011}. A random version of (\ref{imp_tree}) is obtained  by adding random residuals as in (\ref{RI}).  Bagging and boosting methods may be used to improve the efficiency of tree-based procedures. This is discussed next.

\subsection{Random forests}\label{RF}

Random forest \citep{breiman2001random} is an ensemble method which achieves better  accuracy than tree-regression methods by creating a large number of different regression trees and combining them to produce more accurate predictions than a single model would. Random forests are especially efficient in complex settings such as small sample sizes, high-dimensional predictor space and complex relationships (\cite{hamza2005empirical}, \cite{Diaz-Uriarte2006}, among others). Since the article of \cite{breiman2001random}, random forests have been extensively used in various fields such as medicine \citep{Fraiwan2012}, time series analysis \citep{Kane2014}, agriculture \citep{Grimm2008}, to cite just a few. Recently, their theoretical properties have been established by \cite{scornet2015consistency}.


There exist a number of random forest algorithms (see \citet{biau2016random} for discussion). A widely used algorithm proceeds as follows \citep{dagdoug_goga_haziza_RF_NR}:
\begin{enumerate}[Step 1:]
	\item Consider $B$ bootstrap data sets $D_1, D_2, ...,  D_{B},$ obtained by selecting with replacement $n_r$ pairs $(y_i, \bx_i)$ from $D = \left\{(y_i,\bx_i)\right\}_{i \in S_r}$.
	\item In each bootstrap data set $D_b$ for $b=1, \ldots, B,$ fit a regression tree and determine the prediction $\widehat{f}_{tree}^{(b)}$ for the unknown $f$ in (\ref{first_mom}) as described in section \ref{CART}. For each regression tree, only $p'$ variables randomly chosen among the $p$ variables are considered in the  search for the best split in (\ref{opt1}).
	\item  The imputed value for missing $y_i$ is obtained by averaging the predictions at the point $\bx_i$ of the $B$ regression tree predictions:
	\begin{equation}\label{imp_RF}
	\widehat{y}_i = \dfrac{1}{B} \sum_{b = 1}^B \widehat{f}_{tree}^{(b)}(\bx_i), \quad i\in S_m,
	\end{equation}
where $\widehat{f}_{tree}^{(b)}(\bx_i)$ is the prediction for the unknown $f$ in (\ref{first_mom}) computed at $\bx_i$ and obtained with the $b$th regression tree as described in Section \ref{CART}. More specifically, from (\ref{hat_f_tree}), the prediction $\widehat{f}_{tree}^{(b)}(\bx_i)$ corresponds to the weighted average of $y$-values for $j\in S_r$ falling in the same region $A^{(b)}(\bx_i)$ containing $i\in S_m$.
\end{enumerate}
 A random version of (\ref{imp_RF}) is obtained  by adding random residuals as in (\ref{RI}).
Although random forests are based on fully-grown trees,  the accuracy of the predictions is improved by considering  bootstrap of units and model aggregation, a procedure called  \textit{bagging} and used in statistical learning for reducing the variability. The number $B$ of regression trees should be large enough to ensure a good performance without harming the processing time; see \cite{scornet2017tuning}.
The second improvement brought by random forest is the random selection at each split   of $p'$ predictors,
achieving decorrelated trees. The value of $p'$ is typically chosen as $p'\simeq \sqrt{p}$ \citep{hastie_tibshirani_friedman_2011}. In random forest algorithms, a stopping criterion is usually specified so that the algorithm stops once a certain condition (e.g., on the minimum number of units in each final nodes) is met.

\subsection{Least square tree-boosting and other tree-boosting methods}\label{boosting}

As in bagging, boosting \citep{friedman_2001} is a procedure that can be applied to any statistical learning methods for improving the accuracy of model predictions and  is typically used with tree-based methods. While bagging involves the selection of bootstrap samples to create many different predictions, boosting is an iterative method that starts with a weak fit (or learner) and improves it at each step of the algorithm by predicting the residuals of prior models and adding them together to make the final prediction.

To understand how boosting works, consider a regression tree with non-overlapping regions $A_1, \ldots, A_J,$  expressed as
\begin{eqnarray}
T(x,\Theta)=\sum_{j=1}^J\gamma_{j}\mathds{1}(\mathbf{x}_i\in A_{j}).\label{develop_tree}
\end{eqnarray}
The parameter $\Theta=\{\gamma_{j}, A_j\}_{j=1}^J$ is obtained by minimizing
\begin{eqnarray}
\widehat{\Theta}=\mbox{arg}\min_{\Theta}\sum_{j=1}^J\sum_{i:\mathbf{x}_i\in A_j}\mathcal{L}(y_i,\gamma_j)=\mbox{arg}\min_{\Theta}\sum_{i\in S_r}\mathcal{L}(y_i,T(\mathbf{x}_i,\Theta)),\label{opt_global_tree}
\end{eqnarray}
where $\mathcal L$ denotes a loss function; e.g., the quadratic loss function. With the latter, given a region $A_j,$ estimating the constant $\gamma_j$ is usually straightforward as $\widehat{\gamma}_j=\overline{y}_j$ the average the $y$-values belonging to $A_j. $ However, finding the regions $\{A_j\}_{j=1}^J$ and solving (\ref{opt_global_tree}) in a traditional way may prove challenging and computationally intensive as it requires optimizing over all the parameters jointly. To overcome this difficulty, one may use a greedy top-down recursive partitioning algorithm to find $\{A_j\}_{j=1}^J$ as described in Section \ref{CART}. Alternatively, one may split the optimization problem (\ref{opt_global_tree}) into many simple subproblems that can be solved rapidly. Boosting uses the latter and considers that the unknown $f$ has the following additive form:
\begin{eqnarray}
f(x)=\sum_{m=1}^MT(x,\Theta_m),\label{tree_additive}
\end{eqnarray}
where $T(x,\Theta_m)$ for $m=1, \ldots, M$ are trees determined iteratively by using  a forward stagewise procedure \citep{hastie_tibshirani_friedman_2011}: at each step, a new tree is added to the expansion without modifying the coefficients and parameters of trees already added. Each added tree, usually referred to as a weak-learner,  has a small size and slowly improves the estimation of $f$ in areas where it does not perform well.  For the quadratic loss function, after accounting for the survey weights, the algorithm becomes:  \\
\noindent Step 1: Initialize the algorithm with a constant value: $\widehat f_0(\mathbf{x}_i)=0$
and
$$
\widehat{\gamma}_0=\argminA_{\gamma\in \mathbb{R}}\sum_{i \in S_r} w_i(y_i-\gamma)^2=\frac{1}{\sum_{i \in S_r} w_i}\sum_{i \in S_r} w_i y_i.
$$
\noindent Step 2: For $m=1$ to $M$:
\begin{enumerate}
\item [(a)] Given the current model $\widehat f_{m-1},$ fit the regression tree that best predicts the residuals values $y_i-\widehat f_{m-1}(\mathbf{x}_i), i\in S_r$ and get the terminal regions  $(A_{jm})_{j=1}^{J_m}. $

\item [(b)] Given the terminal regions $A_{jm},$ the optimal constants $\widehat{\gamma}_{jm}$ are found as follows:
$$
\widehat{\gamma}_{jm}=\argminA_{\gamma_{jm}}\sum_{i\in S_r:\mathbf{x}_i\in A_{jm}}w_i\mathcal{L}(y_i,\widehat f_{m-1}(\mathbf{x}_i)+\gamma_{jm})=\argminA_{\gamma_{jm}}\sum_{i\in S_r:\mathbf{x}_i\in A_{jm}}w_i(y_i-\widehat f_{m-1}(\mathbf{x}_i)-\gamma_{jm})^2
$$
for $j=1, \ldots, J_m.$
\item [(c)] Update  $\widehat f_m(\mathbf{x}_i)=\widehat f_{m-1}(\mathbf{x}_i)+T(\mathbf{x}_i, \widehat{\Theta}_m)$ where $\widehat{\Theta}_m=\{A_{jm}, \widehat{\gamma}_{jm}\}_{j=1}^{J_m}$ and $T(\mathbf{x}_i, \widehat{\Theta}_m)=\sum_{j=1}^{J_m}\widehat{\gamma}_{jm}\mathds{1}(\mathbf{x}_i\in A_{jm}). $
\end{enumerate}
\noindent Step 3: Output $\widehat f_M(\mathbf{x}_i)$ and get the imputed value
\begin{equation}\label{imp_boost}
	\widehat y_i=\widehat f_M(\mathbf{x}_i).
\end{equation}
A random version of (\ref{imp_boost}) is obtained  by adding random residuals as in (\ref{RI}).
The number $M$ of trees should not be too large and, for better performances, \cite{hastie_tibshirani_friedman_2011} recommend to consider the same number of splits $J_m=J$ at each iteration. The value of $J$ reflects the level of dominant interactions between the $X$-variables. The value $J=2$ (one split) produces boosted models with only main effects without interactions, whereas the value $J=3$ allows for two-variable interactions.  Empirical studies suggest that $J=6$ generally leads to good results.
As in ridge regression, shrinkage is used with tree boosting. In this case, Step 2. (c) of the above algorithm is replaced by a penalized version:
$$
\widehat f_m(\mathbf{x}_i)=\widehat f_{m-1}(\mathbf{x}_i)+\nu T(\mathbf{x}_i, \widehat{\Theta}_m),
$$
where the parameter $\nu\in (0,1),$ called learning rate, is used to penalized large trees; usually $\nu=0.1$ or $0.01.$ Both $M$ and $\nu$ control the performance of the model prediction.

\subsubsection{XGBoost}\label{XGboost}
\cite{chen2016} suggested a scalable end-to-end tree boosting system called XGBoost which is extremely fast. Here, we adapt the algorithm in order to account for the survey weights. Consider again a tree with formal expression given in (\ref{develop_tree}). This tree learning algorithm consists of minimizing the following objective function at the  $m$-th iteration:
\begin{eqnarray}
\widehat{\Theta}_m=\argminA_{\Theta_m}\{\sum_{i\in S_r}w_i\mathcal L(y_i,\widehat f_{m-1}(\mathbf{x}_i)+T(\mathbf{x}_i, \Theta_m))\}+\Omega(T(x, \Theta_m)),
\end{eqnarray}
where the penalty function $\Omega(T(x, \Theta_m))=\gamma J+\frac{\lambda}{2} \sum_{j=1}^J\gamma^2_j$ penalizes large trees in order to avoid overfitting. The search problem is optimized by using a second-order Taylor approximation of $\mathcal L,$ and ignoring the constant term, the new optimization problem reduces to:
 \begin{eqnarray}
\widehat{\Theta}_m=\argminA_{\Theta_m}\sum_{j=1}^J\left[\gamma_j\sum_{i\in S_r:\mathbf{x}_i\in A_j} w_i g_i+\frac{1}{2}\gamma^2_j(\sum_{i\in S_r:\mathbf{x}_i\in A_j}w_i h_i+\lambda)\right]+ \gamma J,\label{new_objectif}
\end{eqnarray}
where $g_i$ and $h_i$ are the first and second-order derivatives of the loss function computed at $\widehat f_{m-1}(\mathbf{x}_i)$. With the quadratic loss function, $g_i=2(\widehat f_{m-1}(\mathbf{x}_i)-y_i)$ and $h_i=2.$ The new objective function from (\ref{new_objectif}) is a second-order polynomial with respect to $\gamma_j$, so the optimal $\gamma_j$ is easily obtained as $\gamma^*_j=-(\sum_{i\in S_r:\mathbf{x}_i\in A_j} w_ig_i)/(\sum_{i\in S_r:\mathbf{x}_i\in A_j}w_ih_i+\lambda),$ leading to the optimal value of the objective function as $-(1/2)\sum_{j=1}^J(\sum_{i\in S_r:\mathbf{x}_i\in A_j} w_ig_i)^2/(\sum_{i\in S_r:\mathbf{x}_i\in A_j}w_i h_i+\lambda)+\gamma J.$ This value is then used next as a decision criterion in a greedy top-down recursive algorithm to find the optimal regions $A_j$ of the $m$-th tree to be added.

\subsubsection{Bayesian additive regression trees (BART)}\label{BART}

Bayesian additive regression trees \citep[BART]{Chipman2010} is similar to boosting in the sense  that the unknown regression function $f$ has an additive form as in (\ref{tree_additive}). While boosting is completely nonparametric, BART makes a Gaussian assumption on the model errors:
$$
y_i=f(\mathbf{x}_i)+\epsilon_i,\quad \epsilon_i\sim \mathcal{N} \left(0, \sigma^2\right),
$$
where $f(x)=\sum_{m=1}^MT(x,\Theta_m)=\sum_{m=1}^MT_m(x,\Gamma_m)$ is assumed to be a sum of tree functions and  $\Gamma_m=\left\{\gamma_j, \gamma_2, \ldots, \gamma_{Jm} \right\}$ is the set of parameter values associated with the $J_m$ terminal nodes in each tree $T(x,\Theta_m)$.


As stated in \cite{Chipman2010}, although similar in spirit to gradient boosting, BART differs from boosting algorithms both by the way it weakens the individual trees by relying on a Bayesian framework, but also on how it performs the iterative fitting. More specifically, a prior is specified for the parameters of the model $(T_1, \Gamma_1), (T_2, \Gamma_2), \ldots, (T_m, \Gamma_m)$ and $\sigma^2$. The prior of $T_m$ can be decomposed into three components :
\begin{enumerate}
	\item The probability that a node at depth $J$ is a terminal node is given by $\alpha\left( 1 + J\right)^{-\beta}$ for $\alpha \in \left(0; 1\right), \  \beta \geq 0.$
	\item The distribution on the splitting variable assignments in each interior node is uniform.
	\item The distribution of the splitting value conditional on the chosen splitting variable is also uniform.
\end{enumerate}

Borrowing the illustrative example of \cite{Chipman2010}, with the parameters $\alpha = 0.95$ and $\beta =2$, trees with $1, 2, 3, 4, 5$ terminal nodes receive prior probabilities of $0.05, 0.55, 0.28, 0.09$ and $0.03$, respectively. Therefore, as in boosting, the BART model tends to favor trees with a small number of terminal nodes. However, the process of restricting the depth of regression trees (or equivalently the number of terminal nodes) in BART is different from the one used in boosting. For boosting, the depth of the trees is fixed by the user and is similar for all trees used in the forest. For BART, the user specifies a probability for the trees to have a certain number of terminal nodes. As a result, the number of terminal nodes is random rather tan fixed. Therefore, it is  likely that trees have only a small number of terminal nodes with the BART model, but this number can vary depending on the data at hand. For $\gamma_j$, a conjugate prior is chosen to make computations simpler; e.g.,   $p (\gamma_{jm} | T_m)$ is assumed to be $\mathcal{N} (\gamma_{\gamma}, \sigma^2_{\gamma})$. Similarly, a conjugate prior is chosen for $\sigma^2$, e.g.,  the inverse chi-square distribution. To generate the posterior distribution, the authors suggest the use of a Gibbs sampler. For general guidelines about the choices of these parameters, see \citet{Chipman2010}. The imputed value for missing $y_i$ is obtained as with the general boosting algorithm given in Section \ref{boosting}, where the prediction of each regression tree is the weighted average of the values in the terminal node containing $\bx_i$.

%

%

\subsection{Cubist algorithm}\label{CUBIST}

Cubist is an updated implementation of the M5 algorithm introduced by \cite{quinlan1992learning} and \cite{quinlan1993combining}. It is an algorithm based on regression trees and linear models, among other ingredients. Initially, Cubist was only available under a commercial license. In 2011, the code was released as open-source. The algorithm proceeds as follows \citep [Chap. 8]{kuhn_johnson}:

\begin{enumerate}[Step 1:]
\item Create a partition $\mathcal{P} = \left\{A_1, A_2, ..., A_T\right\}$ of $\mathbb{R}^p.$ To do so, let $\mathcal{C}_A$ be the set of all possible splits in a node $A$ of cardinality $\ell$, that is, the set of all possible pairs (position, variable). Then, the split is performed using the following criterion:

\begin{align*}
L'(z,j) &= \argmaxA_{(z,j) \in \mathcal{C}_A} \sqrt{ \sum_{ i \in S_r} \left(y_i - \left(\dfrac{1}{n_r}\sum_{j' \in S_r} y_{j'} \right)\right)^2} - \sum_{h = 1}^\ell \dfrac{n_h}{n_r} \sqrt{ \sum_{ i : \bx_i \in D_h} \left(y_i - \left(\dfrac{1}{n_r}\sum_{j': \bx_i \in D_h} y_{j'} \right)\right)^2},
\end{align*}
where $D_1, \ldots, D_\ell$ denote the $\ell$ non-terminal nodes after each of the $\ell-1$ previous splits and $n_h$ denotes the cardinal of elements in the node $D_h$.
\item In each node, a linear model is fitted between the survey variable $Y$ and the auxiliary variables that have been used to split the tree. More specifically, consider the $j$th terminal node $A_j$. Then, there exists a path from the first node to the current node $A_j$ in the graph formed by the tree. This path uses $p'_j$ variables among the set $\{X_1, X_2, ..., X_p\}$. For instance, assume that a partition of $5$ elements is created by the tree shown in Figure \ref{figCub}. Then, the linear model in the node $A_1$ is fitted using the variables that created the path in red, that is, $X_1$, $X_4$ and $X_6$, and so $p'_1 = 3$ for this node. The linear model fitted in the node $A_4$ uses only one variable, $X_1,$ (the green path), so $p'_4 = 1$.
 \begin{figure}[h!]
	\centering
	
	\includegraphics[width=0.9\linewidth]{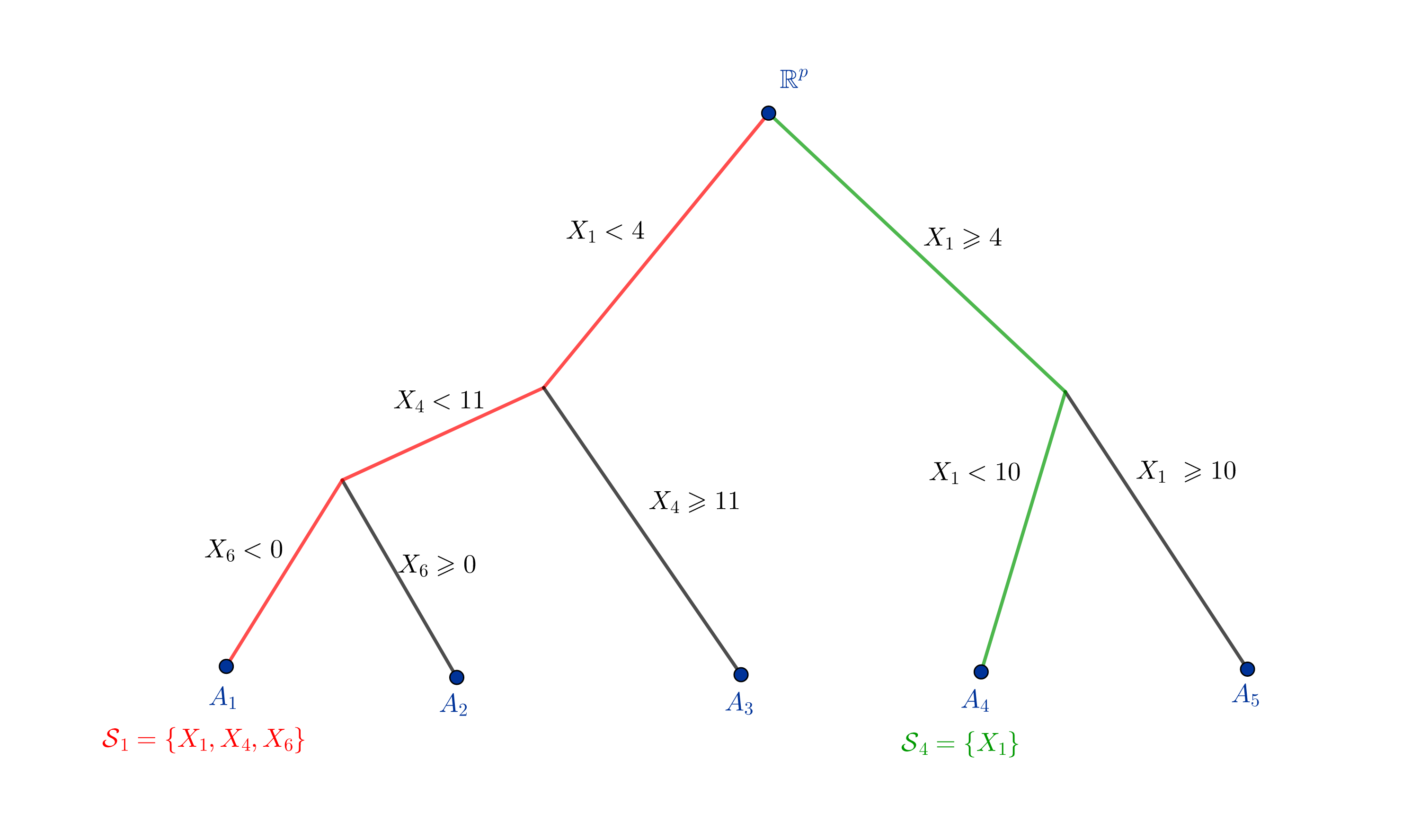}
	
	\caption{Example of a graph induced by a tree algorithm.}
	\label{figCub}			
\end{figure}
The coefficients $\boldsymbol{\beta}_j \in \mathbb{R}^{p'_j}$ of the linear model in the node $A_j$ are estimated  using the customary weighted least squares criterion:
\begin{align*}
\widehat{ \boldsymbol{\beta}}_j  = \argminA_{\boldsymbol{\beta}_j \in \mathbb{R}^{p'_j} }\sum_{i \in S_r} w_i \left\{y_i - \boldsymbol{\beta}_j^\top\bx_i^{(j)} \right\}^2 \mathds{1} \left(\bx_i \in A_j\right),
\end{align*}
where $\bx_i^{(j)}$ is the vector containing the measurements  of the $p'_j$ variables for unit $i$.
\item In each node, a backward elimination procedure is performed using the adjusted error rate (AER) criterion. For instance, in the $j$th terminal node, we have
\begin{equation*}
AER(A_j) = \dfrac{\#(A_j) + p^*}{\#(A_j)- p^*} \sum_{i\in S_r : \bx_i \in A_j} | y_i- \widehat{y}_i|,
\end{equation*}
where $p^*$ denotes the number of variables used in the current model which predicts $\widehat{y}_i$ for a prediction at the point $\bx_i$.
 Each variable  in the initial model is dropped and the AER is recomputed. Terms are dropped from the model as long as the AER decreases.
\item Once the tree is fully grown, it is pruned by removing unnecessary splits. Starting at the terminal nodes, the AER is computed with and without the node. Whenever the node does not result in a decrease of the AER, it is pruned. This process is performed until no more node can be removed.
\item To avoid over-fitting, a smoothing procedure is performed. Let $\widehat{y}_{i(j)}$ be the predicted value obtained by fitting the linear model in the $j$th child node and $\widehat{y}_{i(p)}$ be the predicted value obtained from the direct parent node. These predictions are combined as
\begin{equation*}
\widehat{y}_i = a y_{i(j)} + (1-a)\widehat{y}_{i(p)},
\end{equation*}
where $$a = \dfrac{\widehat{V}(\mathbf{e}_{(p)}) -  \widehat{Cov}(\mathbf{e}_{(j)}, \mathbf{e}_{(p)})}{\widehat{V}(\mathbf{e}_{(j)} - \mathbf{e}_{(p)})}$$
with  $e_{i(j)} = y_i- \widehat{y}_{i(j)}$  denoting the $i$th coordinate of the vector $\mathbf{e}_{(j)}$,  $e_{i(p)} = y_i- \widehat{y}_{i(p)}$ denoting the $i$th coordinate of the vector $\mathbf{e}_{(p)}$ and $\widehat{V}(\cdot)$ and $\widehat{Cov}(\cdot,\cdot)$ denoting the empirical model variance and covariance, respectively.
\item Cubist can be used as an ensemble model. Once the Cubist algorithm is fitted, the subsequent iterations of the algorithm use the previously trained algorithm to define an adjusted response $y_i^{(m)}$ so that the next iteration of the algorithm uses
\begin{equation*}
y_i^{(m)} = y_i - (y_i^{(m-1)} - y_i),
\end{equation*}
where $y_i^{(m)}$ is the value of the adjusted response $y_i$ for the $m$th iteration of the Cubist algorithm.
\item
The final imputed value for missing $y_i$ is derived using a $K$ nearest-neighbour rule:
\begin{equation}\label{imp_Cubist}
\widehat{y}_i = \dfrac{1}{K} \sum_{k= 1}^K \frac{1}{0.5 + d_k} (t_k + \widehat{y}^{(k)} - \widehat{t}_k),
\end{equation}
where  $d_k$ denotes the distance between  $\mathbf{x}_i$ and the $k$th neighbor, $ t_k$ denotes the outcome of the $k$th neighbor and $\widehat{t}_k$ its predicted value.
\end{enumerate}
 A random version version of (\ref{imp_Cubist}) is obtained by adding random residuals as in (\ref{RI}).

\subsection{Support vector regression}\label{SVR}

Support vector machines \citep{Vapnik1998, Vapnik2000, vapnik_cortes_1995, smola2004tutorial} belong to the class of supervised learning algorithms and may be used for regression analysis.
We start by considering the linear regression model
\begin{equation*}
f(\mathbf{x}_i) =  \beta_0 + \mathbf{x}_i^T \boldsymbol{\beta}, \quad \beta_0 \in \mathbb{R}, \quad \boldsymbol{\beta} \in \mathbb{R}^p,
\end{equation*}
before discussing the case of nonlinear relationships. In the customary regression framework, the goal is to minimize the residuals sum of squares. In Support Vector Regression (SVR),  the goal  is to minimize a function of the residuals plus a $L^2$-penalization on the regression coefficient:  
\begin{eqnarray}
\mathcal S=\sum_{i\in S_r}V_{\epsilon}(y_i-f(\mathbf{x}_i))+\frac{\lambda}{2}||\boldsymbol{\beta}||^2,\label{optim_SVR}
\end{eqnarray}
where $V_{\epsilon}$  is the so-called $\epsilon$-insensitive error measure defined as $V_{\epsilon}(x)=0$ if $|x|<\epsilon$ and $|x|-\epsilon$ otherwise \citep{Vapnik2000} for $\epsilon >0;$ $\varepsilon$ can be viewed as the allowed tolerance for fitting; see Figure 1 in \cite{smola2004tutorial}.
The optimization problem (\ref{optim_SVR}) may not have solution and supplementary tolerances $\xi_i, \xi^*_i$ (called also "the slack variables") on the individual  fitted errors are considered  \citep{smola2004tutorial}. There exist several ways for incorporating weights in the optimization problem, leading to different weighted support vector regression solutions.  We consider the method suggested by \cite{lee_song_song_yoon} and \cite{han_clemmensen_2014}:
\begin{eqnarray}
\underset{\beta}{\text{minimize}} \quad \dfrac{1}{2}|| \boldsymbol{\beta} ||^2 + C \sum_{i \in S_r} \widetilde{w}_i \left(\xi_i + \xi_i^*\right)\label{optim_svr2}
\end{eqnarray}
and
\begin{equation}\label{contrainte_svr}
\begin{aligned}
& \text{subject to} & & y_i - \beta_0 - \mathbf{x}_i^T \boldsymbol{\beta} \ \leqslant \epsilon + \xi_i, \\
& &&\beta_0 + \mathbf{x}_i^T\boldsymbol{\beta} -y_i   \ \leqslant \epsilon + \xi_i^*  . \\
&&& \xi_i, \xi_i^* >0,
\end{aligned}
\end{equation}
where $C>0$ is the tuning parameter that provides a trade-off between the smoothness of the fitted function and the deviation from the training data and  $\widetilde{w}_i=w_i/\sum_{j\in S_r}w_j\in (0,1)$ denotes the normalized sampling weight associated with unit $i$. As a result, the $\widetilde{w}_i$'s are all smaller than one. As argued by  \cite{han_clemmensen_2014}, incorporating weights in the objective function as in (\ref{optim_svr2}) has the effect of shrinking the estimators $\widehat{\beta}_j$ to different extents.
The solution of (\ref{optim_SVR}) and (\ref{contrainte_svr}) is given by $\widehat{\boldsymbol{\beta}} = \sum_{ i \in S_r} \left(\widehat{\alpha}_i - \widehat{\alpha}_i^*\right) \mathbf{x}_i,$ which leads to
\begin{equation} \label{learning}
\widehat{f}(\mathbf{x}) = \sum_{ i \in S_r} \left(\widehat{\alpha}_i - \widehat{\alpha}_i^*\right) <\mathbf{x}_i,\mathbf{x}> + \beta_0,
\end{equation}
where $ <\cdot,\cdot>$ is an inner product and $\widehat{\alpha}_i>0$ and $\widehat{\alpha}_i^*>0$ denote the Lagrange multipliers verifying the quadratic programming problem:
$$
\min_{\alpha_i, \alpha^*_i}\epsilon\sum_{i\in S_r}(\alpha_i+\alpha^*_i)-\sum_{i\in S_r}y_i(\alpha_i-\alpha^*_i)+\frac{1}{2}\sum_{i,j\in S_r}(\alpha_i-\alpha^*_i)(\alpha_j-\alpha^*_j)<\mathbf{x}_i,\mathbf{x}_j>
$$
subject to $0\leq \alpha_i, \alpha^*_i\leq  C_i := C \times \widetilde{w}_i,$ $\sum_{i\in S_r}(\alpha_i-\alpha^*_i)=0$ and $\alpha_i\alpha^*_i=0.$
As a result, only a subset of the solution values $(\widehat{\alpha}_i-\widehat{\alpha}^*_i)$ are nonzero and the associated data values are called the support vectors. The solution $\widehat{\boldsymbol{\beta}}$ is written as a linear combination of these support vectors. Moreover, the prediction $\widehat{f}(\mathbf{x})$ uses only the support vectors and the inner products between $\mathbf{x}$ and $\mathbf{x}_i$ without requiring the computation of  $\widehat{\boldsymbol{\beta}}.$ This property is useful for extending the method to handle nonlinear relationships.

We now consider the case of a nonlinear and unknown function $f.$ We approximate $f$ in a basis of functions $\{\phi_m\}_{m=1}^M$ as follows:
$$
f(x)=\sum_{m=1}^M\beta_m\phi_m(x)+\beta_0
$$
and $\beta_0$ and $\boldsymbol{\beta}=(\beta_m)_{m=1}^M$  minimize (\ref{optim_svr2}) and 
\begin{equation}\label{contrainte_svr2}
\begin{aligned}
& \text{subject to} & & y_i - \beta_0 - \sum_{m=1}^M\beta_m\phi_m(x_i) \ \leqslant \epsilon + \xi_i, \\
& &&\beta_0 + \sum_{m=1}^M\beta_m\phi_m(x_i)-y_i   \ \leqslant \epsilon + \xi_i^*  . \\
&&& \xi_i, \xi_i^* >0.
\end{aligned}
\end{equation}
A similar derivation as before leads to
$\widehat{\boldsymbol{\beta}} = \sum_{ i \in S_r} \left(\widehat{\alpha}_i - \widehat{\alpha}_i^*\right) \phi(\mathbf{x}_i)$ for $\phi (\mathbf{x}_i)=(\phi_m(\mathbf{x}_i))_{m=1}^M$
and
\begin{equation*}
\widehat{f}(\mathbf{x}) = \sum_{ i \in S_r} \left(\widehat{\alpha}_i - \widehat{\alpha}_i^*\right) \mathcal{K}(\mathbf{x}_i, \mathbf{x}) + \beta_0,
\end{equation*}
where $\mathcal{K}(\mathbf{x}_i, \mathbf{x})=<\phi(\mathbf{x}_i),\phi(\mathbf{x})>=\sum_{m=1}^M\phi_m(\mathbf{x}_i)\phi_m(\mathbf{x})$ is a positive definite kernel \citep{smola2004tutorial}. The computation of  $\widehat{f}(\mathbf{x})$ involves $\phi(\mathbf{x})$ only through inner products and using a kernel function makes the computation of $\widehat{f}(\mathbf{x})$  possible without requiring $\phi(\mathbf{x}).$ All is needed is the knowledge of $\mathcal K$. Using $\mathcal{K}$, it is possible to solve the optimization problem in a higher-dimensional space without having to compute any product in this space.
 Common choices of $\mathcal{K}(\cdot, \cdot)$ include the Gaussian kernel $\mathcal{K}(\mathbf{x}_i, \mathbf{x}_j) = \exp \left(- ||\mathbf{x}_i - \mathbf{x}_j ||^2\right)$ and the polynomial kernel $\mathcal{K}(\mathbf{x}_i, \mathbf{x}_j) = \left(1 + \mathbf{x}_i^{\top}\mathbf{x}_j\right)^q, \quad q=2, 3, \ldots.$
The imputed value for the missing $y_i$ is given by
\begin{equation}\label{imp_SVR}
\widehat{y}_i =  \sum_{ j \in S_r} \left(\widehat{\alpha}_j - \widehat{\alpha}_j^*\right) \mathcal{K}(\mathbf{x}_j, \mathbf{x}_i) + \widehat{\beta}_0.
\end{equation}
 A random version version of (\ref{imp_SVR}) is obtained by adding random residuals as in (\ref{RI}). The reader is referred to \cite{smola2004tutorial} for a discussion on how to estimate ${\beta}_0.$

\section{Simulation study: the case of population totals}\label{simulation_study}
We conducted an extensive simulation study to investigate the performance of the imputation procedures described in Section \ref{imputation_methods} in terms of bias and efficiency.
\subsection{The setup}\label{small-dimension data}
For each scenario, we repeated $R=5, 000$ iterations of the following process:
\begin{itemize}
\item [(i)] A finite population of size $N=10, 000$ was generated. The population consisted of a survey variable $Y$ and a set of predictors $X_1, \ldots, X_p$.
\item [(ii)] From the finite population generated in Step (i), a sample, of size $n,$ was selected according to a given probability sampling design.
\item [(iii)] In each sample, nonresponse to item $Y$ was generated according to a given nonresponse mechanism.
\item [(iv)] The missing values in each sample were imputed using several imputation procedures.
\end{itemize}
We now give a more in-depth discussion of each of the steps (i)-(iv).

We first generated five predictors $X_1, \ldots, X_5,$ according to the following distributions: $X_1$ followed a normal distribution,
$X_1 \sim \mathcal{N} \left(0,1\right);$ $X_2$ followed a Beta distribution,
$X_2 \sim \text{Beta} \left(3,1\right);$ $X_3$ followed a Gamma distribution, $X_3 \sim 2 \times \text{Gamma} \left(3, 2\right);$ $X_4$ followed a Bernoulli distribution,
 $X_4 \sim \mathcal{B} \left(0.7\right);$
 and $X_5$ followed a multinomial distribution,
 $X_5 \sim \text{Mult} \left(0.4, 0.3, 0.3\right).$
 The predictors $X_1$-$X_3$ were continuous, whereas the predictors $X_4$ and $X_5$ were discrete. The predictors $X_1$-$X_3$ were standardized so as to have a zero mean and a variance equal to one.
%
Given the predictors $X_1$-$X_5,$ we generated the continuous survey variables $Y_1, \ldots, Y_8,$ according to the following models:
\begin{itemize}
	\item  $Y_1 = 2 + 2X_1 + X_2 + 2 X_3 +  \mathcal{N}(0 , 1)$;
	\item $Y_{2} = 2 + 2X_1 + X_2 + 2 X_3 + \text{Pareto} (1, 4)$;
	\item $Y_3 = 2 + X_1 + X_2^2 + X_3 + \mathcal{N}(0 , 1)$;
	\item $Y_4 = 2 + 2X_1 + X_2 + 3 X_3 X_4 + 1.5 \mathds{1}(X_5 = 1) - 2\mathds{1}(X_5 = 2) + \mathcal{N}(0 , 1)$;

	\item $Y_5 = 2 + 5 X_1^3 + 4X_2^2 + X_3 X_4 + 1.5 \mathds{1}(X_5 = 1) - 2\mathds{1}(X_5 = 2)  + \mathcal{N}(0 , 1)$;
	\item $Y_6 = 2 + \left(2 X_1 + X_2 + 2X_3\right)^2 +\mathcal{N}(0 , 1) + \text{Beta} (3,1)$;
	\item $Y_7 = 2 + \left(2X_1 + X_2 + 3 X_3 X_4 + 1.5 \mathds{1}(X_5 = 1) - 2\mathds{1}(X_5 = 2) \right)^2+ \mathcal{N}(0 , 1)$;
	\item $Y_{8} = 4 \cos \left(X_1\right) + \mathcal{N}(0 , 1)$;
\end{itemize}
and the binary survey variables as follows:
\begin{itemize}	
	\item $Y_{9} =\mathds{1}(S_1 > 1/2),$ where
	\begin{align*}
	S_1 &= 0.1 + 0.79\exp\left\{1+0.5  \left(0.75+ 2 X1 + 2X_2+ 2X_3 - X_4 -  X_3X_4 \right. \right. \nonumber
	\\ & \left. \left. + 1.5 \mathds{1}(X_5 = 1) - 2\mathds{1}(X_5= 2)\right)\right\}^{-1};
	\end{align*}
	\item $Y_{10} =\mathds{1}(S_{2} > 1/2),$ where $$S_{2} =  0.55 \times  Q + 0.02 -0.01 X_2^3$$ with
	\begin{align}\label{Q}
	 Q & = \exp \left\{1+0.4 \times \left(6.5 + 2 X_1 + 2 X_2 + 2 X_3 - X_4 -X_3 X_4 \right. \right. \nonumber \\
	 & \left. \left. + 1.5 \mathds{1}(X_5 = 1) - 2\mathds{1}(X_5 = 2)\right)\right\}^{-1}.
	\end{align}
\end{itemize}


For the survey variables $Y_2$ and $Y_6$, note that we have generated errors for non-normal distribution to assess the robustness of the BART procedure that assumes a Gaussian distribution for the errors.

From each population, we selected samples, of (expected) size $n=1, 000,$ according to two sampling designs: (a) simple random sampling without replacement and (b) Poisson sampling with probability proportional to the values of the variable $X_5$; i.e., $\pi_i=1,000 \times  (x_{5i}/\sum_{i \in U}x_{5i})$ for all $i\in U.$ Simple random sampling without replacement was used for estimating the finite population total of the continuous survey variables $Y_1$-$Y_6$ and $Y_8$ and the binary variables $Y_{9}$ and $Y_{10},$ whereas Poisson sampling was used for estimating the totals of the survey variables $Y_4$ and $Y_7.$ 


In each sample, nonresponse to the survey variable $Y_\ell,$ $\ell=1, \ldots, 10,$ was generated according to four nonresponse mechanisms. That is, the response indicators $r_i$ were generated from a Bernoulli distribution with probability $p_{gi},$ $g=1,\ldots, 4,$ where
\begin{align*}
\mbox{ (NR1): }   p_{1i} &= 0.1 + 0.79\exp\left\{1+0.5  \left(0.75+ 2 x_{i1} + 2 x_{i2} \right. \right. \nonumber
\\  & \left. \left. + 2 x_{i3} - x_{i4} - x_{i3}x_{i4} + 1.5 \mathds{1}(x_{i5} = 1) - 2\mathds{1}(x_{i5} = 2)\right)\right\}^{-1}; \\
\mbox{ (NR2): } p_{2i} &= 0.5;\\
\mbox{ (NR3): } p_{3i} &= 0.55 \times  q_i + 0.02 -0.01 x_{i2}^3; \\
\mbox{ (NR4): } p_{4i} &= 0.5 \times q_i + 0.13 - 0.1 \left(\sin(x_{i1}) + \cos(x_{i2})\right);
\end{align*}
where $q_i$ is the $i$th value of $Q$ given by (\ref{Q}). In (NR1)-(NR4), the model parameters were set so as to obtain a response rate of about $50\%$ in each sample.

In each sample, the missing values were imputed according to eleven imputation procedures described in section \ref{imputation_methods}.  Some of the imputation procedures required the specification of some parameters (e.g., regularization parameter, depth of a regression tree, choice of a kernel, etc.). We have included several configurations to assess the impact of these parameters on the performance of these procedures. Based on the different configurations, we ended up with twenty-seven imputation procedures. More specifically, we included the following procedures:
\begin{enumerate}[Procedure 1:]
	\item"LR" : Deterministic linear regression imputation; see Section \ref{linearreg}.\label{proc1}
		\item "MWC$\alpha$"~: Mean imputation within classes, where the number of units in each class was set to  $\alpha \in \{50, 100, 250, 500 \};$ see Section \ref{score_method}. \label{proc2}
	\item"HDWC$\alpha$"~: Random hot-deck imputation within classes, where the number of units in each class was set to  $\alpha \in \{50, 100, 250 \}$; see Section \ref{score_method}. \label{proc3}
	\item "$K$NN"~: $K$-Nearest-Neighbours imputation with $K=1$ and $K=5$ nearest neighbours and the euclidian distance and implemented with the $R$-package \texttt{caret}; see Section \ref{KNN}. \label{proc4}
	\item "AMS$\alpha$"~: Additive models based on cubic $B$-splines with $\alpha$  equidistant interiors knots placed at the $x$-quantiles, where $\alpha \in \{5, 10 \}$ and implemented with the $R$-package \texttt{mgcv}; see Section \ref{Bsplines}. \label{proc5}
	\item"CART"~: Imputation through regression trees with the CART algorithm and  implemented with the $R$-package \texttt{rpart}; see Section \ref{CART}.\label{proc6}
	\item "RF1"~: Imputation through  random forest with $B=1000$ trees, one observation per terminal node and 1 predictor considered for the search in each split. "RF2": Random forest with $B=1000$ trees, 5 observations per terminal node and $\sqrt{p}$ predictors considered for each split, where $p$ is the number of $X$-variables used in the imputation model, in our case $p=5$. "RF3" :  Random forest with $B=1000$ trees, 10 observations per terminal node and $\sqrt{p}$ predictors considered for each split.  Simulations were implemented with the $R$-package \texttt{ranger}; see Section \ref{RF}.\label{proc7}
	\item "XGB1": XGBoost algorithm with $M=50$ trees each one with $J=3$ final splits and a learning rate of $0.1$. "XGB2": XGBoost algorithm with $M=100$ trees with $J=6$ and a learning rate of $0.05$. "XGB3": XGBoost algorithm with $M=250$ trees with $J=10$ and a learning rate of $0.01$. Simulations were implemented with the $R$-package \texttt{xgboost}; see Section \ref{XGboost}. \label{proc8}
	\item"BART" : Imputation through Bayesian additive regression trees. Simulations were implemented with the $R$-package \texttt{bartMachine}; see Section \ref{BART}. \label{proc9}
	\item"CUBIST1": Cubist with one model. "CUBIST2" : Cubist with five models. "CUBIST3" : Cubist with 5 models and unbiased estimation. Simulations were implemented with the $R$-package \texttt{Cubist}; see Section \ref{CUBIST}. \label{proc10}
	\item "SVR1": Support vector regression imputation with a Gaussian kernel and the $\nu$ objective function. "SVR2":  Support vector regression imputation with a polynomial kernel of degree 3 and the $\epsilon$-insensitive objective function. "SVR3": Support vector regression imputation with a Gaussian kernel and  the $\epsilon$-insensitive objective function. "SVR4": Support vector regression imputation with a linear kernel and the $\epsilon$-insensitive objective function. Simulations were implemented with  the $R$-package \texttt{e1071}; see Section \ref{SVR}. \label{proc11}
\end{enumerate}

The imputation procedures used in our simulations were based on an imputation model that included the predictors $X_1, \ldots, X_5,$ without any interaction terms.
Except for random hot-deck imputation (Procedure \ref{proc3}) and nearest-neighbour imputation (Procedure \ref{proc4} with $K=1$),  for the binary variables $Y_{9}$ and $Y_{10}$, note that we have generated zeroes and ones from independent Bernoulli distributions with parameter $\widehat{y}_i,$ where $\widehat{y}_i$ denotes the predicted value associated with unit $i$.   Whenever $\widehat{y}_i <0$, we set it to $\widehat{y}_i = 0$. Similarly, when  $\widehat{y}_i > 1,$ we set it to $\widehat{y}_i = 1$.

As a measure of bias of the imputed estimator $\widehat{t}_{imp}$ given by (\ref{imputed}), we computed the Monte Carlo percent relative bias defined as
\begin{equation}\label{RB_MC}
RB_{MC}(\widehat{t}_{imp})=100 \times \dfrac{1}{R} \sum_{r=1}^R \dfrac{ (\widehat{t}_{imp}^{(r)} - t_y)}{t_y},
\end{equation}
where $\widehat{t}_{imp}^{(r)}$ denotes the imputed estimator  $\widehat{t}_{imp}$ at the $r$th iteration, $r=1, \ldots, 5, 000$.

As a measure of efficiency, we computed the relative of efficiency, using the complete data estimator $\widehat{t}_{\pi}$ given by (\ref{HT}), as the reference. That is,
\begin{equation}\label{RE_MC}
RE_{MC}(\widehat{t}_{imp}) =100 \times \dfrac{ MSE_{MC}(\widehat{t}_{imp}) }{ MSE_{MC}(\widehat{t}_{\pi})},
\end{equation}
where $MSE_{MC}( \widehat{t}_{imp} ) = R^{-1} \sum_{r=1}^R ( \widehat{t}_{imp}^{(r)} - t_y )^2$ and $MSE_{MC}(\widehat{t}_{\pi})$ is defined similarly.

\subsection{Simulation results}

In Section \ref{continu_SRS}, we discuss the simulation results pertaining to the continuous survey variables $Y_1, \ldots, Y_6$ and $Y_8,$ with simple random sampling without replacement. The results for Poisson sampling used in the case of $Y_4$ and $Y_7$ are discussed in Section \ref{continu_PO}. Finally, the case of the binary variables $Y_9$ and $Y_{10},$ whose totals were estimated with simple random sampling without replacement, is discussed in Section \ref{discret}.


\subsubsection{Continuous survey variables and simple random sampling without replacement}\label{continu_SRS}

For simple random sampling without replacement,  for each of the twenty-seven imputation procedures, we had seven survey variables and four nonresponse mechanisms, leading to $27 \times  4 \times 27=756$ sets of simulation results. For ease of presentation, we  present the results in tabular and graphic forms. The displayed statistical analyses were  obtained from  $4\times 7=28$ scenarios obtained by crossing all the nonresponse models and the survey variables.

\begin{landscape}
	\begin{figure}[h!]
		\centering	
		\includegraphics[width=1\linewidth]{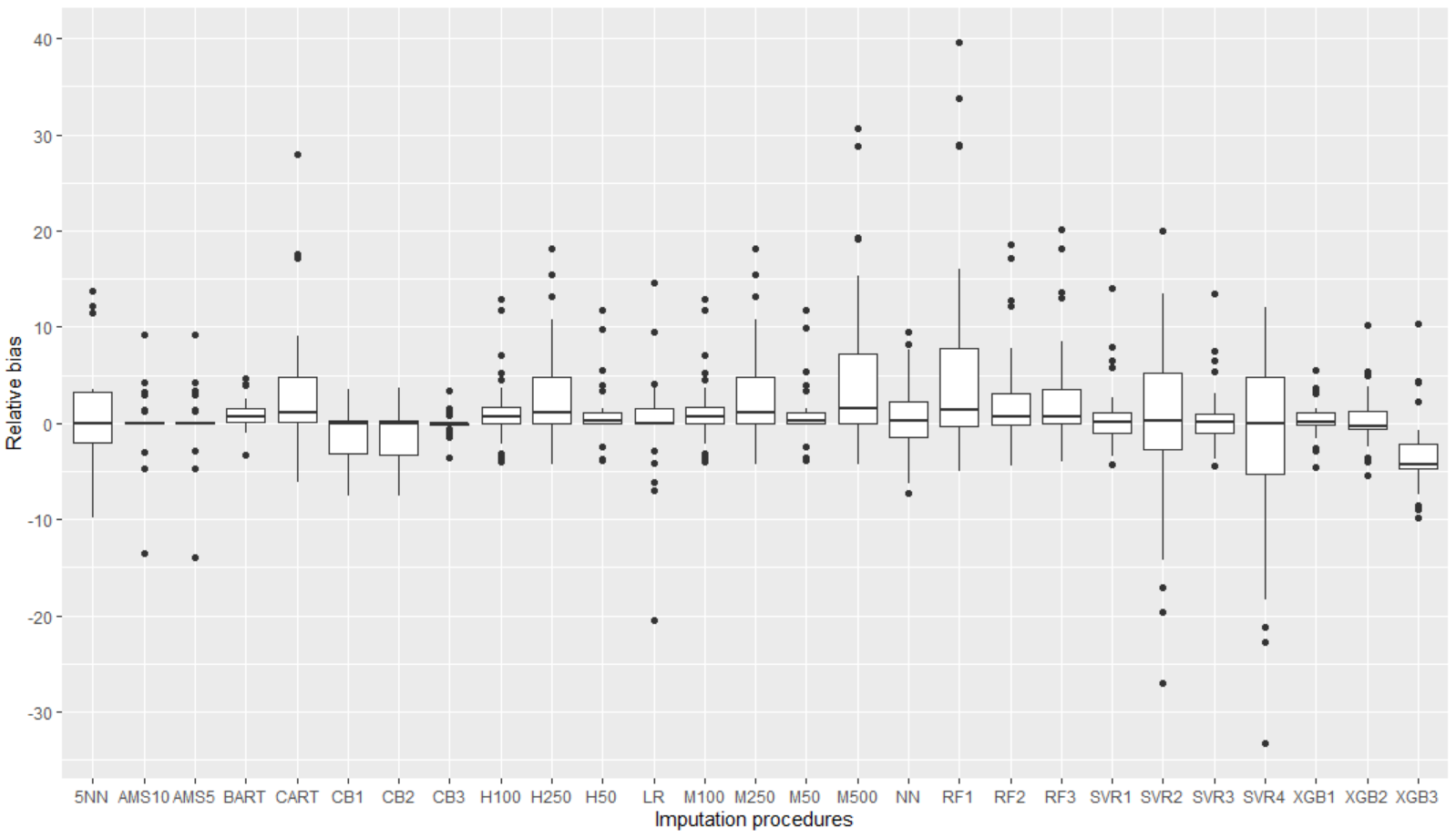}	
		\caption{Monte Carlo percent relative bias across the scenarios.}
		\label{fig_MC_RB}	
	\end{figure}
\end{landscape}
\begin{landscape}
	\begin{figure}[h!]
		\centering
		\includegraphics[width=1\linewidth]{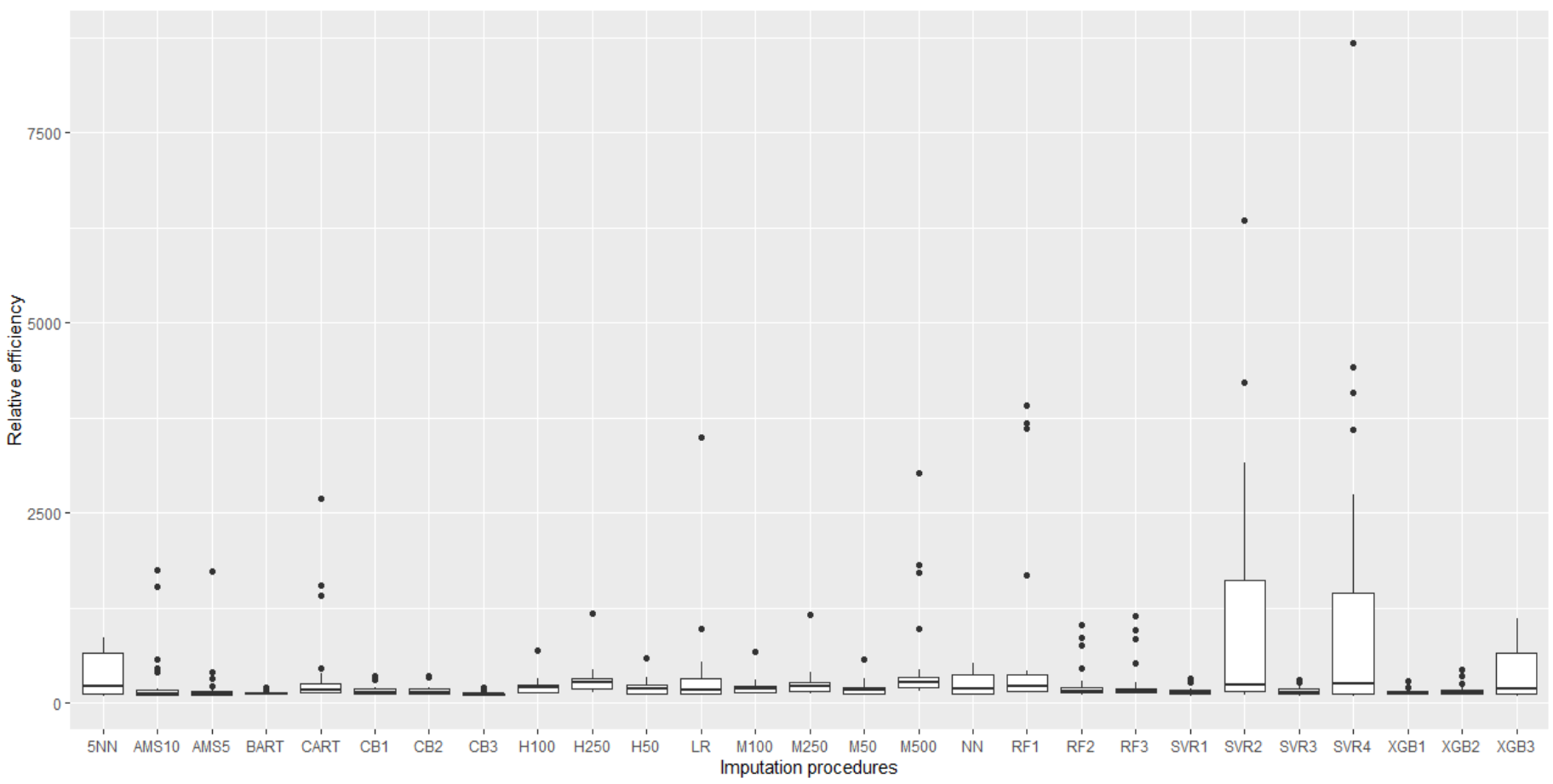}
		\caption{Monte Carlo percent relative efficiency across the scenarios.}
		\label{fig_MC_RE}	
	\end{figure}
\end{landscape}
\begin{figure}[h!]
	\centering
	\includegraphics[width=1\linewidth]{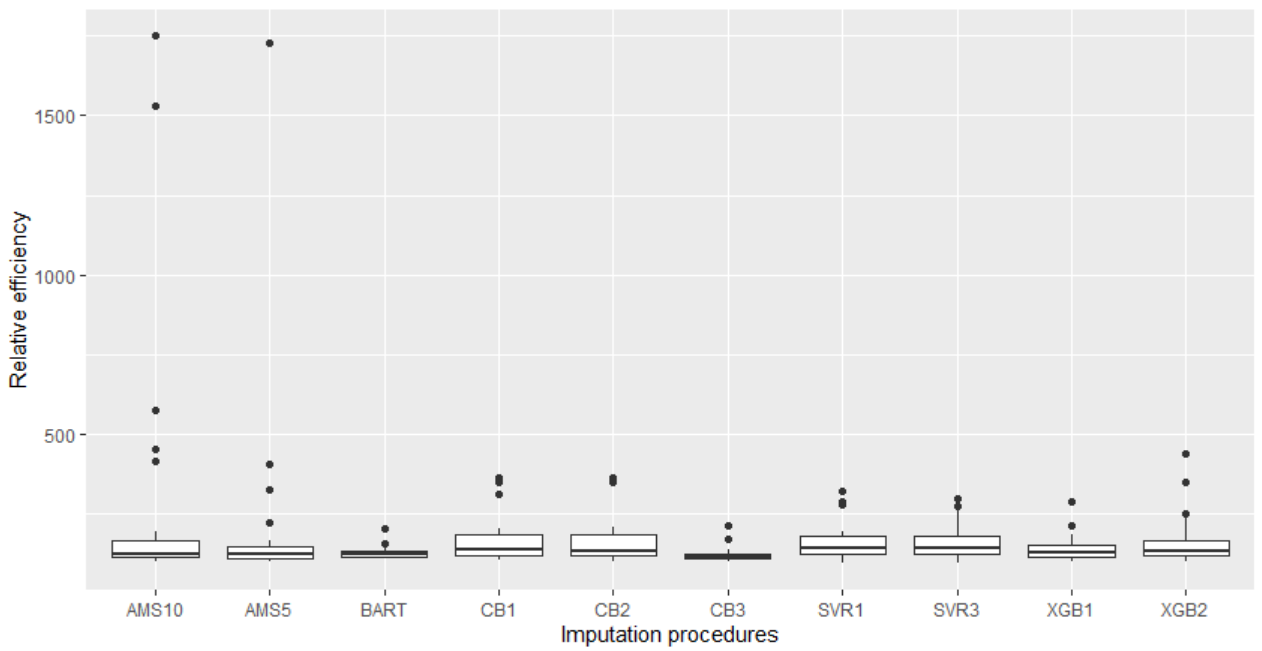}
	\caption{Monte Carlo percent relative efficiency across the scenarios: the best 10 procedures.}
	\label{fig_MC_RE_10}
\end{figure}

For each imputation procedure, Table \ref{tab:RB} and Table \ref{tab:RE} display, respectively, some descriptive statistics regarding the Monte Carlo absolute percent relative bias (absolute value of RB) and the Monte Carlo relative efficiency (RE) of $\widehat{t}_{imp}$ calculated across the twenty-eight scenarios. The corresponding side-by-side boxplots obtained from the twenty-eight scenarios are given in Figures \ref{fig_MC_RB} and \ref{fig_MC_RE}. 
In Tables \ref{tab:RB} and \ref{tab:RE}, the imputation procedures are ordered from the best to the worst with respect to the median absolute percent RB (the median of the twenty-eight values of absolute RB) and the median percent RE (the median of the twenty-eight values of RE), respectively. Figure \ref{fig_MC_RE_10} shows the distribution of the imputed estimator for the best ten imputation procedures in terms of RE. Finally, Table \ref{top5} displays the best five imputation procedures for each $Y$-variable.

%

From Table \ref{tab:RB} and Table \ref{tab:RE}, among the twenty-seven imputation procedures, the best methods were: CUBIST, XGboost, AMS and BART.  The performance of CUBIST3 was especially impressive with a median RE of 115\%, a value of $Q_{95}$ equal to 158\% and a maximum value of 211\%.  The methods XGboost, AMS and BART exhibited similar performances with values of median RE ranging from 122\% and 129\%. However, for some scenarios, these methods did not perform well. For instance, the procedure XGB2 showed a value of max RE of about 438\%, whereas it was equal to 1728\% for AM5. Results suggest that additive models with 5 interiors knots perform better than those with 10 interior knots. The next group of imputation procedures includes SVR and RF, with values of median RE ranging from 141\% and 151\%. Again, for some scenarios, both methods displayed poor performances with values of max RE ranging from 322\% to 1138\%. The procedure CART was less efficient than RF2 and RF3.  The procedure $1$-NN did relatively well with a median RE equal to 194\%. On the other hand, the procedure $5$-NN was rather inefficient with a median RE of 229\%, which suggests that $K$NN with survey data works well only with a small number of neighbour. Turning to mean and random hot-deck imputation within classes, the score method was outperformed by the aforementioned procedures. Among the different versions of MCW and HDWC, the procedure MWC50 (which corresponds to 20 classes) led to the best results. This is consistent with the results of \cite{Haziza2007}. As expected, the procedure HDWC50 was less efficient than MWC50 as random hot-deck imputation suffers from the imputation variance, arising from the random selection of donors within classes. Finally, for some scenarios, it is worth noting that some of the procedures were better than the complete data estimator. For instance, for SVR4, the minimum value of RE and the value of $Q_{0.05}$ were respectively equal to 82\% and 89\%, respectively (see Table \ref{tab:RE}). Finally, the results in Table 5 suggest that the best methods were CUBIST, XGBoost, additive models and BART, which is consistent with the discussion above.

\begin{figure}[h!]
	\centering
	\includegraphics[width=1\linewidth]{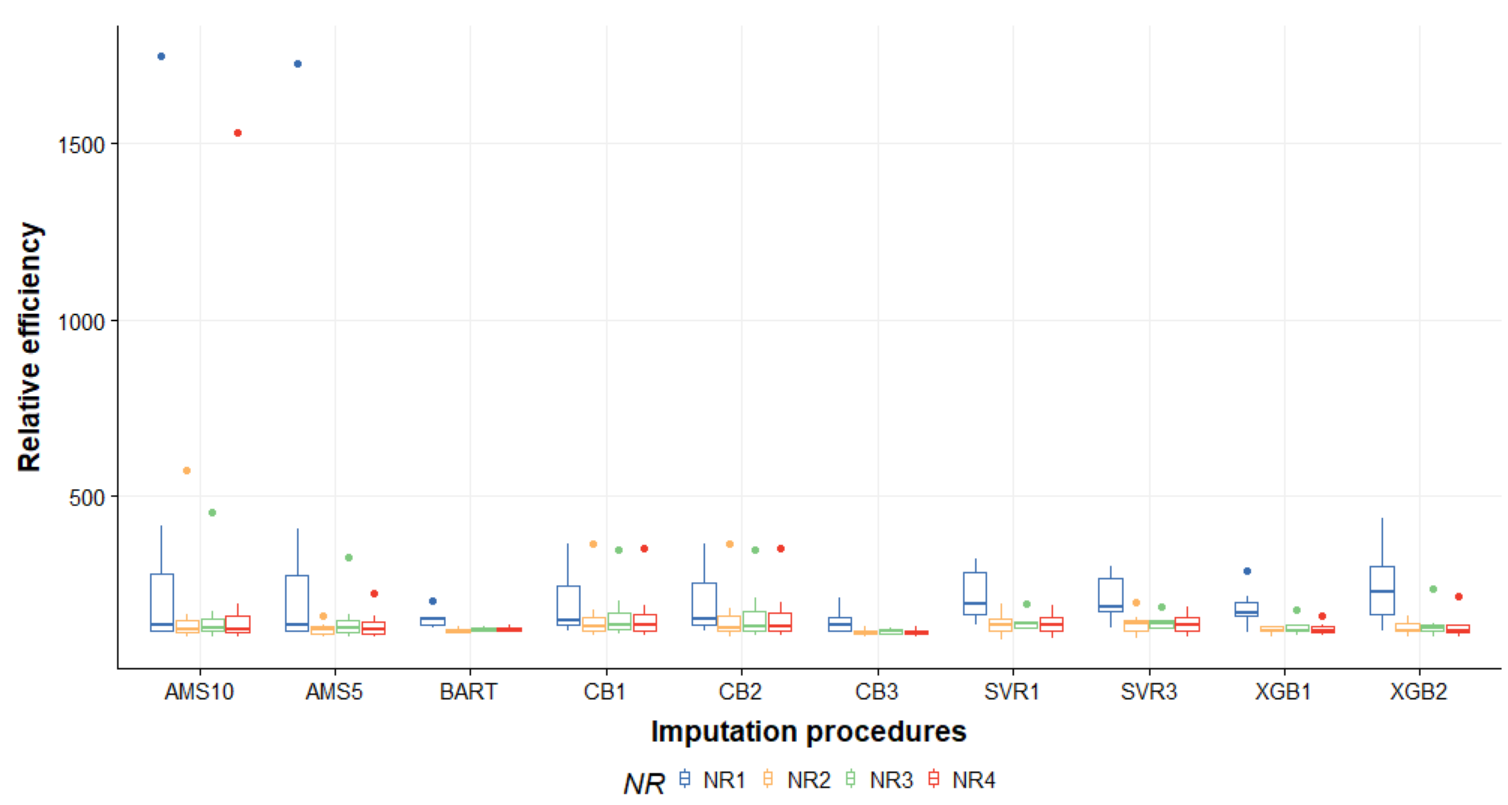}
	\caption[]{The effects of the nonresponse mechanism on the performance of the 10 best imputation procedures.}
	\label{NR_effect}
\end{figure}

For each of the best ten imputation procedures displayed Table \ref{tab:RE}, Figure \ref{NR_effect} displays the distribution of $\widehat{t}_{imp}$ for each nonresponse mechanism. Figure \ref{NR_effect} suggests that the nonresponse mechanism may have a considerable impact on the behavior of the imputed estimator. For instance, in our experiments, we note that most of the imputation procedures performed poorly in the case of the nonresponse mechanism (NR1). Notable exceptions were AMS5, BART and Cubist3. In particular, Cubist3 seemed to be insensitive to the nonresponse mechanism, which is a desirable feature.

%


\begin{table}[]
	\centering
	\resizebox{\textwidth}{!}{%
		\begin{tabular}{lllllllllllllllll}
			\hline
			Ranking &  & Model   &  & Min  &  & $Q_{0.05}$ &  & $Q_{0.25}$ &  & $Q_{0.5}$ &  & $Q_{0.75}$ &  &$Q_{0.95}$ &  & Max   \\ \hline
			&  &         &  &      &  &     &  &      &  &      &  &      &  &      &  &       \\
			1       &  & CUBIST3 &  & 0.0    &  & 0.0   &  & 0.0    &  & 0.1  &  & 0.9  &  & 2.8  &  & 3.5  \\
			2       &  & AMS5    &  & 0.0    &  & 0.0   &  & 0.0    &  & 0.1  &  & 1.8  &  & 7.7  &  & 13.8 \\
			3       &  & AMS10   &  & 0.0    &  & 0.0   &  & 0.0    &  & 0.1  &  & 1.8  &  & 7.6  &  & 13.5 \\
			4       &  & CUBIST1 &  & 0.0    &  & 0.0   &  & 0.1  &  & 0.5  &  & 3.4  &  & 7.5  &  & 7.5  \\
			5       &  & XGB1    &  & 0.0 &  & 0.0   &  & 0.2  &  & 0.6  &  & 1.8  &  & 4.2  &  & 5.4  \\
			6       &  & MWC50   &  & 0.0    &  & 0.0   &  & 0.1  &  & 0.6  &  & 2.7  &  & 8.3  &  & 11.7 \\
			7       &  & HDWC50  &  & 0.0    &  & 0.0   &  & 0.1  &  & 0.6  &  & 2.7  &  & 8.3  &  & 11.8 \\
			8       &  & CUBIST2 &  & 0.0    &  & 0.0   &  & 0.1  &  & 0.6  &  & 3.6  &  & 7.5  &  & 7.5  \\
			9       &  & BART    &  & 0.0 &  & 0.1 &  & 0.4  &  & 0.8  &  & 2.2  &  & 4.0    &  & 4.6  \\
			10      &  & XGB2    &  & 0.1 &  & 0.2 &  & 0.4  &  & 0.9  &  & 2.8  &  & 5.4  &  & 10.1 \\
			11      &  & LR      &  & 0.0    &  & 0.0   &  & 0.1  &  & 0.9  &  & 3.8  &  & 12.8 &  & 20.4 \\
			12      &  & SVR3    &  & 0.1 &  & 0.1 &  & 0.4  &  & 1.0    &  & 3.2  &  & 7.1  &  & 13.5  \\
			13      &  & MWC100  &  & 0.0 &  & 0.0   &  & 0.3  &  & 1.0    &  & 3.6  &  & 10.1 &  & 12.9 \\
			14      &  & HDWC100 &  & 0.0    &  & 0.0   &  & 0.3  &  & 1.0    &  & 3.6  &  & 10.1 &  & 12.9 \\
			15      &  & SVR1    &  & 0.0 &  & 0.1 &  & 0.4  &  & 1.2  &  & 3.4  &  & 7.4  &  & 14.0 \\
			16      &  & RF3     &  & 0.0    &  & 0.2 &  & 0.5  &  & 1.3  &  & 3.8  &  & 16.6 &  & 20.7 \\
			17      &  & RF2     &  & 0.0 &  & 0.1 &  & 0.4  &  & 1.4  &  & 4    &  & 15.6 &  & 18.6  \\
			18      &  & MWC250  &  & 0.0    &  & 0.0   &  & 0.7  &  & 1.7  &  & 4.9  &  & 14.6 &  & 18.1 \\
			19      &  & HDWC250 &  & 0.0 &  & 0.0   &  & 0.6  &  & 1.7  &  & 4.9  &  & 14.6 &  & 18.1 \\
			20      &  & RF1     &  & 0.1 &  & 0.2 &  & 0.9  &  & 1.7  &  & 7.7  &  & 32.1 &  & 39.5 \\
			21      &  & NN      &  & 0.0 &  & 0.1 &  & 1.0    &  & 2.1  &  & 5.2  &  & 8.0    &  & 9.4  \\
			22      &  & MWC500  &  & 0.0    &  & 0.0   &  & 0.7  &  & 2.2  &  & 7.2  &  & 25.5 &  & 30.6 \\
			23      &  & CART    &  & 0.0 &  & 0.1 &  & 0.1  &  & 2.4  &  & 4.9  &  & 17.4 &  & 28.0    \\
			24      &  & X5NN    &  & 0.0 &  & 0.2 &  & 1.5  &  & 3    &  & 7.3  &  & 12.0   &  & 13.7 \\
			25      &  & SVR2    &  & 0.1 &  & 0.2 &  & 1.0    &  & 3.7  &  & 11.7 &  & 19.9 &  & 27.0 \\
			26      &  & XGB3    &  & 0.6 &  & 1.5 &  & 3.1  &  & 4.3  &  & 5.0    &  & 9.5  &  & 10.3 \\
			27      &  & SVR4    &  & 0.0    &  & 0.0   &  & 2.4  &  & 5.3  &  & 7.8  &  & 22.2 &  & 33.3 \\ \hline
		\end{tabular}%
	}
	\caption{Monte Carlo percent absolute relative bias of the imputed estimator:
		Descriptive statistics over all the scenarios}
	\label{tab:RB}
	
\end{table}

\begin{table}[h!]
	\centering
	\resizebox{\textwidth}{!}{%
		\begin{tabular}{lllllllllllllllll}
			\hline
			Ranking &  & Model   &  & Min &  & $Q_{0.05}$   &  & $Q_{0.25}$  &  & $Q_{0.5}$  &  & $Q_{0.75}$   &  & $Q_{0.95}$   &  & Max  \\ \hline
			&  &         &  &     &  &       &  &       &  &       &  &        &  &        &  &      \\
			1       &  & CUBIST3 &  & 102 &  & 102 &  & 111   &  & 115   &  & 125  &  & 158  &  & 211  \\
			2       &  & BART    &  & 113 &  & 113   &  & 116 &  & 122 &  & 131  &  & 154 &  & 204  \\
			3       &  & AMS5    &  & 100 &  & 101   &  & 111 &  & 123   &  & 147  &  & 378  &  & 1728 \\
			4       &  & AMS10   &  & 100 &  & 101   &  & 112 &  & 123 &  & 167    &  & 1195&  & 1749 \\
			5       &  & XGB1    &  & 101 &  & 103 &  & 115 &  & 129 &  & 153  &  & 203  &  & 288  \\
			6       &  & CUBIST2 &  & 102 &  & 103 &  & 119 &  & 133   &  & 187  &  & 360  &  & 365  \\
			7       &  & XGB2    &  & 102 &  & 102   &  & 117   &  & 133 &  & 166 &  & 316  &  & 438  \\
			8       &  & CUBIST1 &  & 103 &  & 105 &  & 120   &  & 136   &  & 182 &  & 360  &  & 365  \\
			9       &  & SVR1    &  & 94  &  & 103&  & 122&  & 141   &  & 180    &  & 284  &  & 322  \\
			10      &  & SVR3    &  & 95  &  & 106 &  & 122 &  & 143   &  & 181  &  & 269    &  & 299  \\
			11      &  & RF3     &  & 115 &  & 118&  & 131 &  & 149   &  & 192  &  & 919 &  & 1138 \\
			12      &  & RF2     &  & 113 &  & 118 &  & 130 &  & 151 &  & 202 &  & 824 &  & 1025 \\
			13      &  & CART    &  & 125 &  & 134&  & 143 &  & 168   &  & 248    &  & 1498 &  & 2683 \\
			14      &  & LR      &  & 110 &  & 111   &  & 114 &  & 169 &  & 315    &  & 823  &  & 3494 \\
			15      &  & MWC50   &  & 113 &  & 114   &  & 122 &  & 171   &  & 205  &  & 308  &  & 583  \\
			16      &  & HDWC50  &  & 120 &  & 120   &  & 128 &  & 189 &  & 240 &  & 332  &  & 600  \\
			17      &  & MWC100  &  & 116 &  & 116 &  & 136   &  & 191   &  & 217  &  & 296  &  & 670  \\
			18      &  & NN      &  & 101 &  & 111 &  & 125 &  & 194   &  & 378  &  & 486  &  & 526  \\
			19      &  & XGB3    &  & 92  &  & 100   &  & 128   &  & 194   &  & 663 &  & 1082 &  & 1104 \\
			20      &  & HDWC100 &  & 123 &  & 125   &  & 142 &  & 213   &  & 246  &  & 322  &  & 686  \\
			21      &  & RF1     &  & 136 &  & 137 &  & 149 &  & 223 &  & 375    &  & 3656 &  & 3916 \\
			22      &  & MWC250  &  & 128 &  & 130 &  & 159&  & 229 &  & 279    &  & 383  &  & 1162 \\
			23      &  & 5NN    &  & 94  &  & 108&  & 123 &  & 229 &  & 659  &  & 775  &  & 855  \\
			24      &  & SVR2    &  & 97  &  & 102 &  & 151 &  & 242 &  & 1616 &  & 3849 &  & 6355 \\
			25      &  & SVR4    &  & 82  &  & 89  &  & 117   &  & 258 &  & 1439 &  & 4301 &  & 8675 \\
			26      &  & HDWC250 &  & 141 &  & 143&  & 185 &  & 265   &  & 325    &  & 411  &  & 1184 \\
			27      &  & MWC500  &  & 151 &  & 155 &  & 202   &  & 269   &  & 336  &  & 1783   &  & 3021 \\ \hline
		\end{tabular}%
	}
	\caption{Monte Carlo percent absolute relative efficiency of the imputed estimator:
		Descriptive statistics over all the scenarios}
	\label{tab:RE}
\end{table}

%

%

%

\begin{table}[h!]
	\centering
	\resizebox{\textwidth}{!}{%
		\begin{tabular}{lllllllllllllllll}
			\hline
			Ranking &  & Y1   && Y2   &  & Y3      &  & Y4         &  & Y5  && Y6  &  & Y7      &  & Y8      \\ \hline
			&  &         &  &   &&      &  &         &  &               &  &    &&     &  &         \\
			1       &  & LR   && CUBIST3   &  & AMS5 &  & BART        &  & XGB3 && CUBIST3 &  & CUBIST3 &  & CUBIST3 \\
			&  &    &&     &  &         &  &         &  &                &  &      &&   &  &         \\
			2       &  & CUBIST3 &  & LR &&AMS10    &  & CUBIST3    &  & AMS5 &&BART &  & AMS5    &  & AMS5    \\
			&  &         &  &    &&     &  &         &  &         &  &               &  &      &&   \\
			3       &  & MW50  &&AMS5  &  &BART    &  & CUBIST1     &  & AMS10 && SVR3&  & AMS10   &  & AMS10   \\
			&  &         &  &    &&     &  &         &  &               &  &         &  &    &&     \\
			4       &  & AMS5&&MWC50   &  & CUBIST3    &  & CUBIST2  &  & XGB1 &&SVR1  &  &MWC50     &  &   XGB1 \\
			&  &         &  &   &&      &  &         &  &         &  &               &  &        && \\
			5       &  & AMS10&&AMS10   &  & CUBIST2    &  & XGB1     &  & XGB2&& XGB1  &  & BART    &  & BART    \\ \hline
		\end{tabular}%
	}
	\caption{Best 5 imputation procedures for each survey variable.}
	\label{top5}
\end{table}

\subsubsection{Continuous survey variables with Poisson sampling}\label{continu_PO}
Recall that Poisson sampling was used for estimating the population total of the survey variables $Y_4$ and $Y_7$.  This led to $2 \times 4\times 27=216$ sets of results. Due to the small number of scenarios ($ 2 \times 4 =8$) for each of the survey variables $Y_4$ and $Y_7$,  Tables \ref{PO1} and \ref{PO2} show the minimum, the median and the maximum Monte Carlo  percent absolute RB and Monte Carlo percent RE only. The size variable $X_5$ used to obtain the first-order inclusion probabilities was included as a predictor in the imputation models. The results in Tables \ref{PO1} and \ref{PO2}  were consistent with those obtained for simple random sampling without replacement. Again, the best methods  were  CUBIST3, BART and XGB1 in terms of either bias or efficiency.
\begin{table}[h!]
	\centering
	\resizebox{0.55\textwidth}{!}{%
		\begin{tabular}{lllllllll}
			\hline
			Ranking   &  & Model   &  & Min &  & $Q_{0.5}$ &  & Max \\ \hline
			&  &           &  &        &  &      &  &        \\
			1  &  & BART    &  & 0.1   &  & 0.9  &  & 3.0   \\
			2  &  & CUBIST3 &  & 0.0   &  & 1    &  & 6.5   \\
			3  &  & XGB1    &  & 0.0  &  & 2.4  &  & 5.2   \\
			4  &  & CUBIST1 &  & 0.0  &  & 3.4  &  & 10.9 \\
			5  &  & RF2     &  & 0.3   &  & 3.5  &  & 15.8  \\
			6  &  & RF3     &  & 0.5   &  & 3.5  &  & 16.8  \\
			7  &  & XGB2    &  & 0.4   &  & 3.9  &  & 8.6   \\
			8  &  & AMS5    &  & 0.2   &  & 4.3  &  & 11.1  \\
			9  &  & AMS10   &  & 0.2  &  & 4.3  &  & 10.7  \\
			10 &  & CUBIST2 &  & 0.0   &  & 4.3  &  & 12.6  \\
			11 &  & RF1     &  & 0.8    &  & 4.4  &  & 31.4   \\
			12 &  & SVR3    &  & 0.1   &  & 4.4  &  & 6.7   \\
			13 &  & LR      &  & 0.2   &  & 4.9  &  & 16.8  \\
			14 &  & SVR1    &  & 0.1    &  & 4.9  &  & 7.1   \\
			15 &  & MWC500  &  & 0.0   &  & 5.0    &  & 26.1  \\
			16 &  & NN      &  & 0.0      &  & 5.0    &  & 7.3   \\
			17 &  & MWC250  &  & 0.0   &  & 5.1  &  & 14.7  \\
			18 &  & HDWC50 &  & 0.8   &  & 5.1  &  & 9.9   \\
			19 &  & MWC50   &  & 0.0      &  & 5.2  &  & 9.9   \\
			20 &  & MWC100  &  & 0.0      &  & 5.2  &  & 10.1   \\
			21 &  & HDWC100 &  & 0.1    &  & 5.2  &  & 10.0  \\
			22 &  & HDWC250 &  & 0.0   &  & 5.2  &  & 14.7  \\
			23 &  & CART    &  & 0.2   &  & 5.6  &  & 24.6  \\
			24 &  & 5NN     &  & 1.3   &  & 7.1  &  & 11.7  \\
			25 &  & XGB3    &  & 2.5   &  & 8.8  &  & 11.1  \\
			26 &  & SVR2    &  & 1.0   &  & 11.7 &  & 22.6 \\
			27 &  & SVR4    &  & 0.2   &  & 15.4 &  & 27.5 \\ \hline
		\end{tabular}%
	}
	\caption{Monte Carlo percent absolute relative bias of the imputed estimator:
		Descriptive statistics for Poisson sampling.}
		\label{PO1}
\end{table}

\begin{table}[h!]
	\centering
	\resizebox{0.55\textwidth}{!}{%
		\begin{tabular}{lllllllll}
			\hline
			Ranking  &  & Model   &  & Min &  & $Q_{0.5}$  &  & Max \\ \hline
			&  &           &  &        &  &       &  &        \\
			1  &  & BART    &  & 106    &  & 117 &  & 139    \\
			2  &  & CUBIST3 &  & 111    &  & 118 &  & 239    \\
			3 &  & XGB1    &  & 108    &  & 133 &  & 207    \\
			4  &  & RF2     &  & 114    &  & 144 &  & 565    \\
			5  &  & RF3     &  & 114    &  & 145 &  & 621    \\
			6  &  & XGB2    &  & 110    &  & 156   &  & 246    \\
			7  &  & SVR3    &  & 109    &  & 165 &  & 198    \\
			8  &  & AMS5    &  & 124    &  & 168 &  & 486    \\
			9  &  & SVR1    &  & 109    &  & 175   &  & 209    \\
			10 &  & CUBIST1 &  & 114    &  & 175   &  & 469    \\
			11 &  & NN      &  & 117    &  & 178   &  & 234    \\
			12 &  & MWC50   &  & 125    &  & 188 &  & 396    \\
			13 &  & MWC100  &  & 125    &  & 188 &  & 363    \\
			14 &  & RF1     &  & 122    &  & 188 &  & 1868   \\
			15 &  & LR      &  & 123    &  & 189   &  & 923    \\
			16 &  & MWC250  &  & 128    &  & 190   &  & 525    \\
			17 &  & CUBIST2 &  & 111    &  & 193 &  & 548    \\
			18 &  & CART    &  & 133    &  & 198   &  & 1224   \\
			19 &  & MWC500  &  & 133    &  & 198 &  & 1346   \\
			20 &  & HDWC50  &  & 135    &  & 210   &  & 409    \\
			21 &  & HDWC100 &  & 139    &  & 213 &  & 381    \\
			22 &  & HDWC250 &  & 145    &  & 217 &  & 539    \\
			23 &  & 5NN     &  & 120    &  & 241 &  & 370    \\
			24 &  & XGB3    &  & 116    &  & 272 &  & 441    \\
			25 &  & AMS10   &  & 130    &  & 313 &  & 592    \\
			26 &  & SVR2    &  & 142    &  & 493 &  & 1619   \\
			27 &  & SVR4    &  & 141    &  & 769   &  & 2119  \\ \hline
		\end{tabular}%
	}
\caption{Monte Carlo percent relative efficiency of the imputed estimator:
	Descriptive statistics for Poisson sampling.}
	\label{PO2}
\end{table}


\subsubsection{Binary survey variables}\label{discret}

In this section, we present the results pertaining to the binary variables $Y_9$ and $Y_{10}$. Again, for each imputation procedure, we obtained  $2 \times 4=8$ sets of results. Tables \ref{BI1} and \ref{BI2} show the minimum, the median and the maximum Monte Carlo percent absolute RB and Monte Carlo percent RE, respectively.

The ranking for binary survey variables was slightly different from that obtained for the continuous survey variables. Nearest-neighbor (NN) imputation procedure was the best in terms of bias and efficiency. Recall that NN imputation did not rank among the best procedures for the continuous variables. NN imputation was followed by CUBIST, XGBOOST and BART.

\begin{table}[h!]
	\centering
	\resizebox{0.55\textwidth}{!}{%
		\begin{tabular}{lllllllll}
			\hline
			Ranking &  & Model   &  & Min &  & $Q_{0.5}$ &  & Max \\ \hline
			&  &           &  &        &  &       &  &        \\
			1     &  & NN      &  & 136    &  & 144 &  & 428    \\
			2     &  & XGB3    &  & 153    &  & 165 &  & 860    \\
			3     &  & XGB2    &  & 156    &  & 167   &  & 827    \\
			4     &  & CUBIST3 &  & 156    &  & 167 &  & 841    \\
			5     &  & XGB1    &  & 156    &  & 171 &  & 932    \\
			6     &  & BART    &  & 156    &  & 173 &  & 1052   \\
			7     &  & 5NN     &  & 152    &  & 174 &  & 1191   \\
			8     &  & CUBIST2 &  & 163    &  & 179 &  & 873    \\
			9    &  & CUBIST1 &  & 169    &  & 191   &  & 904    \\
			10    &  & RF2     &  & 158    &  & 192   &  & 1572   \\
			11    &  & RF3     &  & 162    &  & 198   &  & 1769   \\
			12    &  & AMS5    &  & 169    &  & 219 &  & 2453   \\
			13    &  & MWC100    &  & 160    &  & 221 &  & 1120   \\
			14    &  & MWC50   &  & 159    &  & 222   &  & 1067   \\
			15    &  & SVR1    &  & 171    &  & 222   &  & 3196   \\
			16    &  & AMS10   &  & 165    &  & 223   &  & 2472   \\
			17    &  & MWC50     &  & 159    &  & 223 &  & 1061   \\
			8    &  & M100  &  & 159    &  & 225   &  & 1116   \\
			19    &  & CART    &  & 176    &  & 229   &  & 1882   \\
			20    &  & LR      &  & 164    &  & 230 &  & 2707   \\
			21    &  & MWC250  &  & 172    &  & 244 &  & 1460   \\
			22    &  & MWC250    &  & 173    &  & 246 &  & 1471   \\
			23    &  & SVR3    &  & 191    &  & 280   &  & 2899   \\
			24    &  & RF1     &  & 190    &  & 305   &  & 4666   \\
			25    &  & M500    &  & 186    &  & 365 &  & 4977   \\
			26    &  & SVR4    &  & 219    &  & 409 &  & 26429  \\
			27    &  & SVR2    &  & 413    &  & 1839  &  & 17279 \\ \hline
		\end{tabular}%
	}
	\caption{Monte Carlo percent relative efficiency of the imputed estimator:
		Descriptive statistics for the binary survey variables.}\label{BI1}
\end{table}

\begin{table}[h!]
	\centering
	\resizebox{0.55\textwidth}{!}{%
		\begin{tabular}{lllllllll}
			\hline
			Ranking &  & Model   &  &  Min &  & $Q_{0.5}$ &  & Max \\ \hline
			&  &           &  &        &  &      &  &        \\
			1    &  & NN     &  & 0.0   &  & 0.5  &  & 3.6   \\
			2     &  & CUBIST3 &  & 0.02   &  & 0.7  &  & 6.7  \\
			3     &  & XGB3    &  & 0.03   &  & 0.8  &  & 7.7   \\
			4    &  & BART    &  & 0.1    &  & 0.8  &  & 8.8   \\
			5    &  & XGB1    &  & 0.14   &  & 0.9  &  & 7.9   \\
			6     &  & XGB2    &  & 0.0      &  & 0.9  &  & 6.9  \\
			7     &  & 5NN    &  & 0.0   &  & 1.0    &  & 7.3   \\
			8     &  & CUBIST2 &  & 0.2    &  & 1.0    &  & 7.0   \\
			9     &  & CUBIST1 &  & 0.0      &  & 1.1  &  & 6.8   \\
			10    &  & RF2     &  & 0.12   &  & 1.5  &  & 10.3   \\
			11    &  & RF3     &  & 0.13   &  & 1.6  &  & 11.0  \\
			12    &  & AMS5    &  & 0.04   &  & 1.6  &  & 11.9  \\
			13    &  & AMS10   &  & 0.1   &  & 1.6  &  & 11.9  \\
			14    &  & SVR1    &  & 0.3    &  & 1.7  &  & 12.0  \\
			15    &  & LR      &  & 0.19   &  & 1.8  &  & 12.3  \\
			16    &  & CART    &  & 0.18   &  & 1.8  &  & 11.4  \\
			17    &  & MWC50   &  & 0.0      &  & 1.8  &  & 7.5   \\
			18    &  & MWC100  &  & 0.0      &  & 1.8  &  & 7.7   \\
			19    &  & HDWC50  &  & 0.03   &  & 1.8  &  & 7.5   \\
			20    &  & HDWC100 &  & 0.01   &  & 1.8  &  & 7.7   \\
			21    &  & MWC250  &  & 0.0      &  & 2.0    &  & 9.4   \\
			22    &  & HDWC250 &  & 0.0      &  & 2.0    &  & 9.4   \\
			23    &  & SVR3    &  & 0.43   &  & 2.3  &  & 11.5  \\
			24    &  & RF1     &  & 0.08   &  & 2.7  &  & 19.0   \\
			25    &  & SVR4    &  & 0.17   &  & 3.0    &  & 36.5  \\
			26    &  & MWC500  &  & 0.0   &  & 3.2  &  & 16.4  \\
			27    &  & SVR2    &  & 1.9    &  & 9.5  &  & 33.9 \\ \hline
		\end{tabular}%
	}
	\caption{Monte Carlo percent absolute relative bias of the imputed estimator:
		Descriptive statistics for the binary survey variables.}\label{BI2}
\end{table}

\subsection{High-dimensional setting}
In this section, we investigate the performance of a subset of the imputation procedures considered in Section \ref{small-dimension data} in a high-dimensional setting.  To that end, we used data from the Irish Commission for Energy Regulation (CER) Smart Metering Project conducted in 2009-2010 (CER, 2011) that focused on energy consumption and energy regulation\footnote{The data are available on request at: \texttt{https://www.ucd.ie/issda/data/commissionforenergyregulationcer/}. }.
About 6000 smart meters were installed in Irish residences and businesses. The customer's electrical consumption was collected every half an hour over a period of about two years.

We considered a subset of the original data set. We ended up with a population of $N=6291$ smart meters (households and businesses) for a period of $14$ consecutive days. For each population unit $i$ (household or business), we had $2 \times 7 \times 48 = 672$ measurements denoted by $X_j=X(t_j), j=1, \ldots 672$. Each of these 672 measurements represents the electricity consumption (in kW) at instant $t_j$. We denote by $x_{ij}$ the value of $X_j$ recorded by the smart meter $i$ for $i=1, \ldots, N$ at instant $t_j$.
It should be noted that these variables were highly correlated among themselves with a condition number of the matrix $N^{-1}\mathbf{X}^T\mathbf{X}$ computed using all the data, of about $60.000.$


We created four survey variables based on a subset of the auxiliary variables $X_1, \ldots, X_{672}$:
\begin{align*}
	Y_1 &= 400+2X_1 +X_2 +2X_3 +\mathcal{N}(0, 1500); \\
	Y_2 &=400+X_1 X_2 +2 X_3 +\mathcal{N}(0, 1500);\\
	Y_3 &=500 + 2 X_4 + 400 \mathds{1}_{\{X_5 > 156\}} - 400 \mathds{1} \left(X_5 \leqslant 156\right) + 1000 \mathds{1} \left(X_2 > 190\right) \\&+300 \mathds{1}\left(X_5 > 200\right) + \mathcal{N}(0, 1500);\\
	Y_4 &= 1 + \cos(2X_1 + X_2 + 2X_3)^2 + \epsilon_1,
\end{align*}
where $\epsilon_1 \sim \mathcal{E}(2)$ and these error terms were centered so as to have a  mean equal to zero. We were interested in estimating the population total of the survey variables $Y_1$-$Y_4$. Again, the simulation was based of $R=5, 000$ iterations of the process described in Section \ref{simulation_study}. Samples of size $n = 1000$ were selected according to  simple random sampling without replacement. The missing values to the survey variables $Y_1$-$Y_4$ were generated according to
\begin{equation*}
p_i = 0.1 + 0.89\times \text{sigmoid}  \left\{-0.83 + 0.001\times(2x_{i1} + 2x_{i2} - 2.5x_{i3}) \right\},
\end{equation*}
leading to an average response rate of about $50\%$.

Three high and very high dimensional settings were considered: in the first setting, the imputation models used  the first $15$ auxiliary variables $X_1, ..., X_{15},$ in the data set.  In the second and third settings, the imputation models were based on the first $100$ and $300$ auxiliary variables $X_1, ..., X_{100},$ and $X_1, ..., X_{300},$ respectively.

To impute the missing values, we confined to a subset of the imputation procedures considered in Section \ref{small-dimension data}: additive models,  BART, CUBIST, XGBoost, random forests, nearest-neighbour imputation and support vector regression. Linear regression imputation and mean imputation within  20 classes were also considered. It is well known that the quality of predictions based on linear models tend to deteriorate substantially in the presence of a very large number of auxiliary variables. To cope with this issue, we also considered  principal components analysis as a reduction-dimension method;  see \cite{cardot_goga_shehzad_2017}.

Table \ref{tab101}  shows the Monte Carlo percent relative bias (RB) and relative efficiency (RE)  for $p=15$ predictors.   Table \ref{tab4} shows the results for  $p=100$ and $p=300$ predictors. For each scenario, the best imputation procedures are highlighted in bold. Note that the relative efficiency is now computed with respect to the mean square error of the imputed estimator based on the true imputation model. The additive models were considered in the first setting only ($p=15$ variables) because their performance deteriorated rapidly as the number $p$ of variables increased. For $p=100$ and $p=300$ the backfitting algorithm did not reach convergence in most scenarios.

From Tables \ref{tab101} and  \ref{tab4}, we note that CUBIST and XGBoost were the best method in the vast majority of the scenarios.  These methods were followed by BART and random forests. As expected, additive models performed poorly, which illustrates the curse of dimensionality. It is worth pointing out that random forests performed better in the high-dimensional setting than they did in the low-dimension setting considered in section \ref{small-dimension data}. Finally, the strategy based on principal components analysis did relatively well in most scenarios.

\begin{table}[h!]
	\centering
	\resizebox{\textwidth}{!}{%
		\begin{tabular}{llllllllllllll}
			\hline
			Variable    & Criterion & LR     & MWC50 & RF2  & XGB1  & NN    & SVR3  & AMS5  & CB3 & PCR1   & PCR2   & PCR3   & BART \\ \hline
			&           &        &       &      &       &       &       &       &         &        &        &        &      \\
			\multirow{2}{*}{$Y_1$} & RE        & \textbf{100}    & 117   & 110  & \textbf{103}   & 111   & 124   & \textbf{101}   & \textbf{100}     & 160    & 113    & \textbf{100}    & \textbf{101}  \\
			& RB        & -0,18  & 1,7  & 1,7& 0     & -0,1 & 2,6  & -0,0 & -0,1   & 4,0   & 0,6   & -0,5  & 0,3  \\
			&           &        &       &      &       &       &       &       &         &        &        &        &      \\
			\multirow{2}{*}{$Y_2$} & RE        & 184    & 176   & \textbf{103}  & \textbf{100}   & \textbf{100}   & 295   & 7041  & \textbf{101}     & 159    & 213    & 207    & 106  \\
			& RB        & -44,3 & 15,7 & 3,8 & 0     & 0,7  & 19,2 & 9,5  & -0,0   & -47,0 & -53,1 & -48,5 & 2,1 \\
			&           &        &       &      &       &       &       &       &         &        &        &        &      \\
			\multirow{2}{*}{$Y_3$} & RE        & 190    & 135   & \textbf{102}  & 108   & 128   & 134   & 403   & 109     & 188    & 178    & 210    & \textbf{105}  \\
			& RB        & 4,6   & 2,1  & 0,1 & -0,2 & 0,1  & 2,08  & -0,0 & 1,2     & 4,6   & 4,3   & 5,2   & 0,0 \\
			&           &        &       &      &       &       &       &       &         &        &        &        &      \\
			\multirow{2}{*}{$Y_4$} & RE        & 125    & 126   & 143  & 147   & 188   & 195   & 130   & \textbf{118}     & \textbf{119}    & 121    & 123    & 131  \\
			& RB        & -0,0  & -0,0 & 0,5 & 0,2  & -0,1 & -1,3 & 0,0     & -0,0   & -0,11  & -0,1  & -0,0  & 0,0 \\ \hline
		\end{tabular}%
	}
	\caption{Relative biais (RB) and relative efficiency (RE) of imputation procedures with $p=15$ auxiliary variables.}
	\label{tab101}
\end{table}


\begin{table}[h!]
	\centering
	\resizebox{\textwidth}{!}{%
		\begin{tabular}{llllllllllllll}
			\hline
			Variable              & Dim                    & Criterion & LR     & MWC50 & RF2  & XGB1  & NN   & SVR3  & CB3   & PCR1   & PCR2   & PCR3   & BART  \\ \hline
			&                        &           &        &       &      &       &      &       &       &        &        &        &       \\
			\multirow{2}{*}{$Y_1$}   & \multirow{2}{*}{p=100} & RE        & \textbf{102}    & 122   & 149  & \textbf{103}   & 216  & 187   & \textbf{100}   & 269    & 226    & 151    & \textbf{105}   \\
			&                        & RB        & 0,14   & 2,1  & 4,2 & 0,3  & 6,2 & 5,1  & 0     & 7,8    & 6,6   & 4,0   & 0,6  \\
			&                        &           &        &       &      &       &      &       &       &        &        &        &       \\
			\multirow{2}{*}{$Y_2$}   & \multirow{2}{*}{p=100} & RE        & 115    & 287   & \textbf{109}  & \textbf{100}   & \textbf{100}  & 340   & \textbf{100}   & \textbf{100}    & \textbf{108}    & 140    & 127   \\
			&                        & RB        & -23,8 & 34,3 & 7,5  & 0,1  & 3,3 & 26,1 & -0,0 & -31,0 & -28,9 & -32,5 & 5,8   \\
			&                        &           &        &       &      &       &      &       &       &        &        &        &       \\
			\multirow{2}{*}{$Y_3$}   & \multirow{2}{*}{p=100} & RE        & 158    & 185   & \textbf{107}  & \textbf{107}   & 354  & 162   & \textbf{108}   & 236    & 224    & 196    & 129   \\
			&                        & RB        & 3,2    & 3,9  & 1,1 & -0,0 & 7,0 & 3,4  & 0,9  & 5,9   & 5,5   & 4,8   & 7,7  \\
			&                        &           &        &       &      &       &      &       &       &        &        &        &       \\
			\multirow{2}{*}{$Y_4$}   & \multirow{2}{*}{p=100} & RE        & 140    & 141   & 151  & 146   & 243  & 217   & \textbf{122}   & \textbf{120}    & \textbf{120}    & \textbf{121}    & 135   \\
			&                        & RB        & 0,0   & 0,1  & 0,7 & 0,28  & 0,4 & -1,5 & -0,0 & -0,0  & -0,1   & -0,1  & -0,0 \\
			&                        &           &        &       &      &       &      &       &       &        &        &        &       \\
			\multirow{2}{*}{$Y_1$}   & \multirow{2}{*}{p=300} & RE        & 120    & 215   & 190  & 103   & 286  & 237   & \textbf{100}   & 290    & 262    & 189    & \textbf{110}   \\
			&                        & RB        & -0,2  & 1     & 5,7 & 0,6  & 7,05 & 6,7  & 0,06  & 8,3   & 7,7   & 5,7   & 1,3  \\
			&                        &           &        &       &      &       &      &       &       &        &        &        &       \\
			\multirow{2}{*}{$Y_2$}   & \multirow{2}{*}{p=300} & RE        & \textbf{102}    & 1106  & 112  & \textbf{100}   & \textbf{100}  & 405   & \textbf{100}   & \textbf{91}     & \textbf{85}     & 109    & 243   \\
			&                        & RB        & -6,3  & 89,1  & 9,5 & 0,1  & 4,01 & 35,  & -0,0 & -28,4 & -25,3 & -26,9 & 4,6  \\
			&                        &           &        &       &      &       &      &       &       &        &        &        &       \\
			\multirow{2}{*}{$Y_3$}   & \multirow{2}{*}{p=300} & RE        & 197    & 378   & 118  & 107   & 630  & 180   & \textbf{108}   & 350    & 245    & 224    & 242   \\
			&                        & RB        & 1,0   & 6,7  & 2,0 & 0,0  & 9,1 & 4,1  & 0,8  & 6,2   & 6,1   & 5,6   & 6,4   \\
			&                        &           &        &       &      &       &      &       &       &        &        &        &       \\
			\multirow{2}{*}{$Y_4$} & \multirow{2}{*}{p=300} & RE        & 276    & 584   & 155  & 143   & 443  & 214   & \textbf{124}   & \textbf{120}    & \textbf{120}    & \textbf{121}    & 131   \\
			&                        & RB        & 0,1  & 2,4 & 0,7 & 0,3  & 0,6 & -1,5& 0,06  & -0,0  & -0,1   & -0,1  & -0,0 \\ \hline
		\end{tabular}%
	}
	\caption{Relative biais (RB) and relative efficiency (RE) of imputation procedures with $p=100$ and respectively, $p=300$ auxiliary variables.}
	\label{tab4}
\end{table}

\section{Simulation study: the case of population quantiles}\label{estimation_quantiles}

In this section, we turn our attention to population quantiles. Except for nearest-neighbour imputation, we confined to the random versions of the imputation procedures described in Section \ref{imputation_methods}.
The target parameters were the quantiles of order $\gamma_1 = 0.25$, $\gamma_2 = 0.5$ and $\gamma_3 = 0.75$ that correspond to the first quartile, the median and the third quartile, respectively. We considered a subset of the scenarios described in Section \ref{small-dimension data}. First, we confined to the case of the survey variables $Y_3$ and $Y_6$  and the nonresponse mechanisms (NR1) and (NR3) described in Section \ref{small-dimension data}, leading to $2 \times 2=4$ scenarios. Also, samples were selected according to simple random sampling without replacement only. In each sample, we computed the imputed estimator $	\widehat{\mathcal{Q}}_{\gamma, imp}$ given by (\ref{imputed_quantile}) for $\gamma_1 = 0.25$, $\gamma_2 = 0.5$ and $\gamma_3 = 0.75$. As in Section \ref{simulation_study}, we computed the Monte Carlo percent relative bias of $	\widehat{\mathcal{Q}}_{\gamma, imp}$ and the relative efficiency, given respectively by (\ref{RB_MC}) and (\ref{RE_MC}) with $\widehat{t}_{imp}$ replaced with $	\widehat{\mathcal{Q}}_{\gamma, imp},$ $\widehat{t}_{\pi}$ replaced with $	\widehat{\mathcal{Q}}_{\gamma}$  and $t_y$ replaced with ${Q}_{\gamma}.$

\begin{figure}[h!]
	\centering
	\includegraphics[width=1\linewidth]{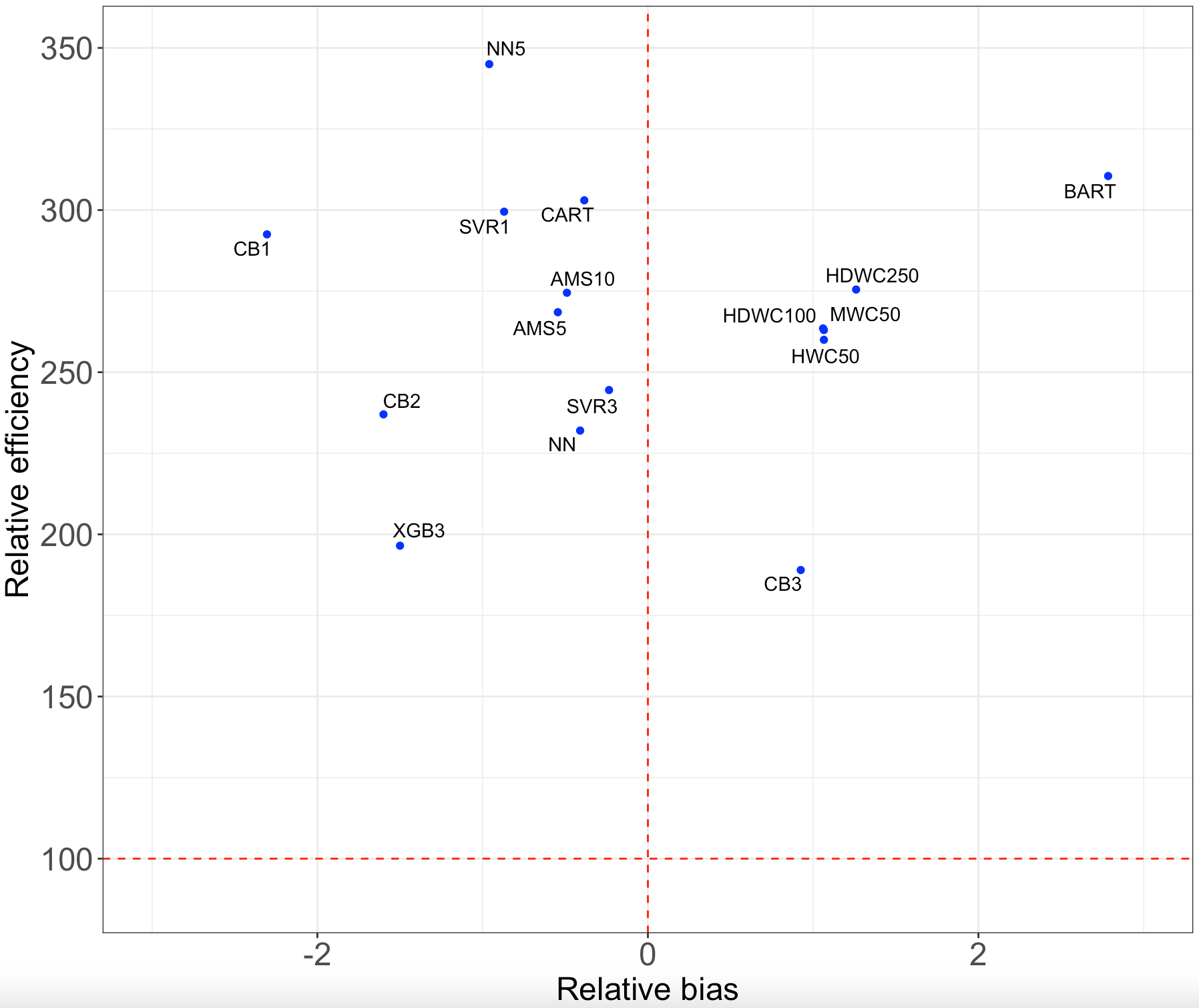}
	\caption[]{Median performances of the best imputed estimators for the estimation of $\mathcal{Q}_{0.25}$.}
	\label{Q1plot}
\end{figure}

\begin{figure}[h!]
	\centering
	\includegraphics[width=1\linewidth]{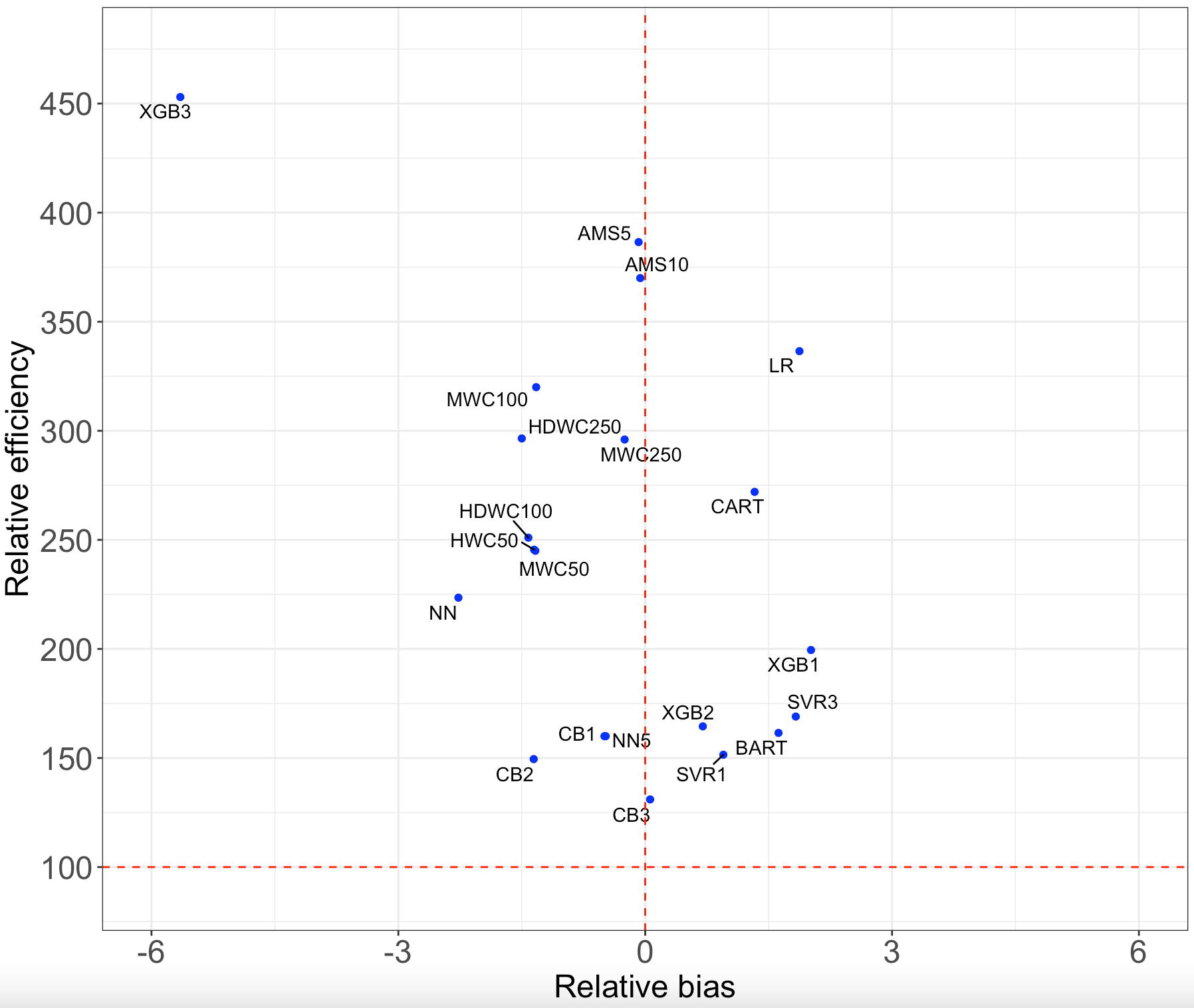}
	\caption[]{Median performances of the best imputed estimators for the estimation of $\mathcal{Q}_{0.5}$.}
	\label{Q2plot}
\end{figure}

\begin{figure}[h!]
	\centering
	\includegraphics[width=1\linewidth]{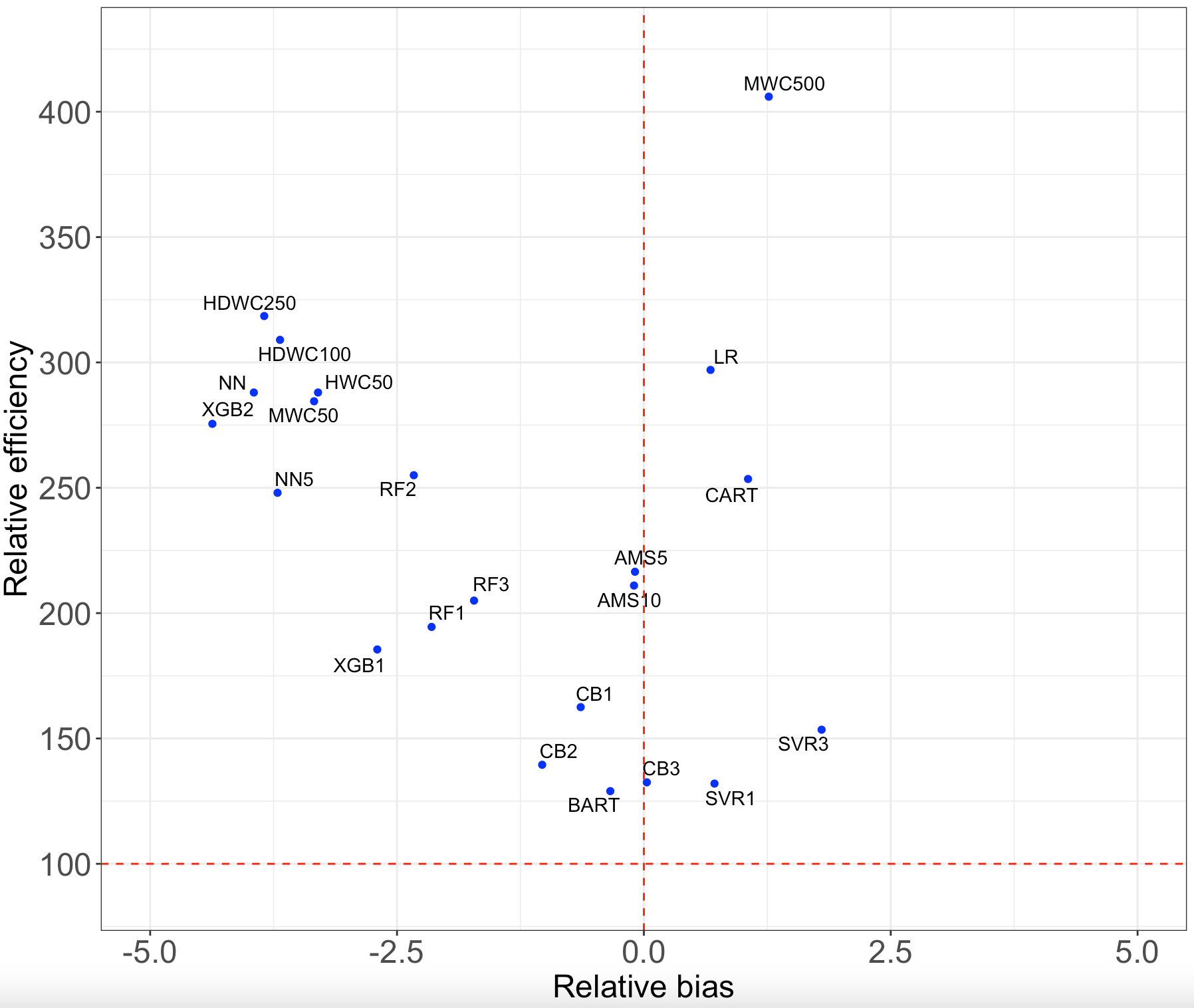}
	\caption[]{Median performances of the best imputed estimators for the estimation of $\mathcal{Q}_{0.75}$.}
	\label{Q3plot}
\end{figure}

The results are presented in Figures \ref{Q1plot}-\ref{Q3plot}. In each figure, the $x$-axis corresponds to the median of the Monte Carlo percent relative bias of  $	\widehat{\mathcal{Q}}_{\gamma, imp}$ computed across the 4 scenarios, whereas the $y$-axis corresponds to the median of the Monte Carlo relative efficiency. For the purpose of clarity, we have excluded from Figures \ref{Q1plot}-\ref{Q3plot} any imputation procedure whose  median of the Monte Carlo percent relative bias lied outside the interval $\left[-20; 20\right]$  or whose median  of the Monte Carlo relative efficiency was above 500.


From Figures \ref{Q1plot}-\ref{Q3plot},  Cubist displayed a very good performance in terms of bias and efficiency for the three quantiles. The procedure XGBoost led to good results for ${Q}_{0.25}$ and ${Q}_{0.75}$ but performed poorly for ${Q}_{0.5}.$ Similarly, BART performed very well for both ${Q}_{0.5}$ and ${Q}_{0.75}$ but exhibited a poor performance for  ${Q}_{0.25}.$ Support vector machine (SVR3) did relatively well for both ${Q}_{0.5}$ and ${Q}_{0.75}$ but was outperformed by Cubist and XGBoost for ${Q}_{0.25}.$ Again, the Cubist algorithm seemed to be insensitive to the target parameter, the model that has generated the $Y$-variable and the nonresponse mechanism, at least in our experiments.

%
%


%

\section{Final remarks}
In this paper, we have conducted an extensive simulation study to compare several nonparametric and machine learning imputation procedures in terms of bias and efficiency. The imputation procedures were evaluated in the case of finite population totals of continuous and binary variables and for population quantiles under both simple random sampling without replacement and proportional-to-size Poisson sampling.
The Cubist algorithm, BART and XGBoost performed very well in a wide variety of settings. In general, these methods seem to be highly robust to model misspecification and seem to have the ability to capture nonlinear trends in the data. Additive models based on $B$-splines performed well in the case of population totals when the number of explanatory variables was small but broke down for large values of $p$.  Finally, random forests performed relatively well in a high-dimensional setting. In practice, the choice of an imputation procedure is not clear-cut and depends on the data at hand. If one is reasonably confident about the correct specification of the first moment of the imputation model (that includes the correct specification of the functional form and the correct specification of the vector of explanatory variables), parametric imputation procedures are expected to do well in terms of bias and efficiency. In addition, parametric imputation is simpler to understand and the results are easier to interpret, in general. In the case of complex/nonlinear relationships and/or in a high-dimensional setting, our empirical investigations suggest that machine learning procedures outperform traditional imputation procedures as they tend to be robust against model misspecification.  However, these procedures require the specification of some regularization parameters. For instance, for XGBoost, one must specify the learning rate, the maximal depth and the coefficient of penalization. In support vector regression, the cost function and the kernel function must be selected, among others. In practice, the value for some of these parameters are determined through a cross-validation procedure. To keep the processing time at a reasonable level,  all the regularization parameters were predetermined in our experiments. Overall, it seems that Cubist is an excellent choice as it performed well in all the scenarios, unlike its main competitors (e.g., XGBoost, BART, random forest, etc.) whose performance varied from one scenario to another.  From a computational point of view, most procedures were efficient. One notable exception is BART that proved to be highly computer intensive with an average processing time approximately twenty times larger than what was required for the other procedures.

Drawing inferences from survey data requires a variance estimate. It is well known that imputed values should not be treated as observed values. Otherwise, the resulting variance estimates tend to be much smaller, on average, than the true variance, especially if the nonresponse rates are appreciable. In the last three decades, a number of variance estimation procedures have been proposed for obtaining variance estimates that account for sampling, nonresponse and imputation. The reader is referred to \cite{hazizaVariance2020} for a comprehensive overview of variance estimation procedures in the presence of singly imputed data sets. Estimating the variance of imputed estimators obtained through machine learning procedures is challenging and requires further research. If the sampling fraction is negligible, one can recourse to the bootstrap procedure of \cite{shao1996bootstrap} that consists of selecting bootstrap samples according to a complete data bootstrap procedure and reimputing the missing values within each bootstrap sample using the same imputation method that was used on the original data. If a machine learning procedure is used to impute the missing data, the Shao-Sitter procedure may be highly computer intensive. When the sampling fraction is not negligible, the problem of bootstrap variance estimation is more intricate \citep{chen2019pseudo}.  To make the variance estimation process simpler for survey practitioners, it would be desirable to derive a "universal" variance estimator based on Taylor expansion procedures that could be applicable to a wide class of machine learning imputation procedures, at least in the case of negligible sampling fractions. This is currently under investigation.

Investigating the performance of deep learning methods in the context of imputation for missing survey data would constitute a promising direction for future research. There exist a wide class of deep learning procedures based on relatively sophisticated algorithms that proved to be extremely efficient in the context of unstructured data such as signal processing or text analysis. However, for deep learning procedures to "shine" in terms of efficiency typically requires a huge volume of unstructured data, which is seldom the case in surveys.
In practice, most data sets in surveys consist of structured data and contains, at most, a few millions observations and a few hundred survey variables. As noted by  \cite{choley_2018}:\\ 
 ``(...) \textit{gradient boosting (such as XGBoost) is used for problems where structure data is available, whereas deep learning is used for perceptual problems such as image classification}''. \\
 We believe that the class of imputation procedures considered in this article, that includes bagging and boosting among others, offers a number of very good options that may be applicable to virtually all the surveys conducted by NSOs.

\bibliographystyle{apalike}
\bibliography{biblio}

\addcontentsline{toc}{section}{References}

\end{document}